\documentclass{jfm}
\pdfoutput=1
\usepackage{graphicx}
\usepackage{natbib}
\usepackage{color}
\usepackage{amsmath}
\usepackage{amssymb}
\ifCUPmtlplainloaded \else
  \checkfont{eurm10}
  \iffontfound
    \IfFileExists{upmath.sty}
      {\typeout{^^JFound AMS Euler Roman fonts on the system,
                   using the 'upmath' package.^^J}%
       \usepackage{upmath}}
      {\typeout{^^JFound AMS Euler Roman fonts on the system, but you
                   dont seem to have the}%
       \typeout{'upmath' package installed. JFM.cls can take advantage
                 of these fonts,^^Jif you use 'upmath' package.^^J}%
      }
  \else
  \fi
\fi

\ifCUPmtlplainloaded \else
  \checkfont{msam10}
  \iffontfound
    \IfFileExists{amssymb.sty}
      {\typeout{^^JFound AMS Symbol fonts on the system, using the
                'amssymb' package.^^J}%
       \usepackage{amssymb}%
       \let\le=\leqslant  
       \let\ge=\geqslant  
      }{}
  \fi
\fi

\ifCUPmtlplainloaded \else
  \IfFileExists{amsbsy.sty}
    {\typeout{^^JFound the 'amsbsy' package on the system, using it.^^J}%
     \usepackage{amsbsy}}
    {\providecommand\boldsymbol[1]{\mbox{\boldmath $##1$}}}
\fi

\newsavebox{\astrutbox}
\sbox{\astrutbox}{\rule[-5pt]{0pt}{20pt}}

\usepackage{xcolor}

\newcommand{\red}[1]{{\color{black} #1}}
\definecolor{grey}{gray}{0.5}
\definecolor{fuchsia}{rgb}{255,0,255}
\newcommand{\lb}[1]{{\color{black} #1}}

\newcommand{\fpgs}[1]{{\color{black} #1}}

\title[Turbulent channel flow of dense suspensions]{Turbulent channel flow of dense suspensions of neutrally-buoyant
spheres}

\author[F. Picano, W.-P. Breugem and L. Brandt]%
{
Francesco Picano$^{1,2}$\thanks{Email address for correspondence: francesco.picano@unipd.it},
\ns Wim-Paul Breugem$^3$ \ns and Luca Brandt$^1$}
\affiliation{
$^1$SeRC (Swedish e-Science Research Centre) and Linn\'e FLOW Centre,
\\ KTH Mechanics, SE-100 44 Stockholm, Sweden\\
$^2$ Department of Industrial Engineering,
University of Padova,\\ Via Venezia 1, 35131 Padua, Italy\\
$^3$ Aero and Hydrodynamics Laboratory, Delft University of Technology, Leeghwaterstraat 21,
NL-2628 CA Delft,
The Netherlands
}

\pubyear{2010}
\volume{650}
\pagerange{119--126}
\date{?; revised ?; accepted ?. - To be entered by editorial office}
\begin{document}

\maketitle

\begin{abstract}
Dense particle suspensions are widely encountered in many applications and in 
environmental flows. While many previous studies investigate their rheological properties
in laminar flows, little is known on the behaviour of these suspensions in the turbulent/inertial regime.   
The present study aims to fill this gap by investigating the turbulent flow of a Newtonian fluid laden with 
solid neutrally-buoyant spheres at relatively high volume fractions in a plane channel. 
\fpgs{Direct Numerical Simulation are performed in the range of volume fractions $\Phi=0-0.2$ with
an Immersed Boundary Method used to account for the dispersed phase.}
The results show that the mean velocity
profiles are  significantly altered by the presence of a solid phase with a decrease of the von K{\'a}rm{\'a}n constant in the log-law.  
The overall drag is found to increase with the volume fraction, more than one would expect just considering the increase of the system  viscosity due to the presence of the particles.  
At the highest volume fraction here investigated, $\Phi=0.2$ , the velocity fluctuation intensities and the Reynolds shear stress are found to 
decrease. The analysis of the mean momentum balance shows that the particle-induced stresses govern the dynamics at 
high $\Phi$ and 
are the main responsible of the overall drag increase. In the dense limit, we therefore find a decrease of the turbulence activity
and a growth of the particle induced stress, where the latter dominates for the Reynolds numbers considered here. 
\end{abstract}


\section{Introduction}

Suspensions of solid particles in liquid flows are widely encountered in 
industrial application and environmental problems. 
Sediment transport, avalanches, slurries, pyroclastic flows, oil industry and pharmaceutical processes represent
 typical examples where a step forward in the understanding and modelling of these complex fluids is essential. 
Given the high flow rates typically encountered 
in these applications, inertia strongly influences the flow regime that may be chaotic and turbulent. The main aim of the present 
work is therefore to investigate the interactions between the phases of a suspension in the turbulent regime.   

Suspensions are often constituted by a Newtonian liquid laden with solid particles that may differ for size, shape, density and 
stiffness. Even restricting our analysis to mono-disperse rigid neutrally-buoyant spheres, the laminar flow of these suspension shows peculiar 
rheological properties, such as high effective viscosities, normal stress differences, shear thinning or thickening, and jamming at high volume fractions, 
see e.g.~\cite{stipow_arfm05,wagbra_pt09,mor_ra09} for recent reviews on the
topic. In particular, still dealing with simple laminar flows,  the suspended phase alters the response of the complex fluid 
to the local deformation rate leading, for example, to an increase of the effective viscosity of the
suspension $\mu_e$ with respect to that of the pure fluid $\mu$~\citep{guamor_book}. 
A first attempt to characterise this effect can be traced back to \cite{einstein1906neue,einstein1911berichtigung}
 who provided a linear estimate of the effective viscosity $\mu_e=\mu\,(1+2.5\,\Phi)$,
with $\Phi$ the volume fraction, valid in the dilute regime. 
Few decades later, \cite{batchelor1970stress}  and \cite{batgre_jfm72} derived and proposed a quadratic correction that 
partially accounts for the mutual interactions among particles, which become more and more critical when increasing the volume fraction. 
Indeed, 
the suspension viscosity increases by more than one order of magnitude in the dense regime, until the system jams behaving as a glass or a crystal~\citep{sierou2002rheology}.
For dense cases only semi-empirical laws exist for the effective viscosity; 
the mixture viscosity has been observed to diverge when the system approaches the maximum packing limit $\Phi_m=0.58-0.62$ \citep*{boy_etal_prl12}, 
as reproduced by empirical fits as those by  Eilers and Kriegher \& Dougherty~\citep{stipow_arfm05}.

The rheological properties of suspensions have been often studied in the viscous Stokesian regime where inertial effects are
negligible and can be safely neglected. Nonetheless in several applications the flow Reynolds number is high enough that 
the inertia is significant at the particle scale. 
The seminal work of \cite{Bagnold} on the highly inertial regime revealed how the increase of the particle collisions 
induces an effective viscosity that increases linearly with the shear rate.  
Even if the macroscopic flow is viscous and laminar, inertial effects at the particle scale
may induce shear-thickening~\citep{kulmor_pof08,pic_etal_prl13} or normal stress differences~\citep*{zarraga_etal_jor00}.
This change of the macroscopic behaviour is due to a strong modification of the particle microstructure, i.e.\ the
relative position and velocity of the suspended particles~\citep{mor_ra09,pic_etal_prl13}. A finite particle-scale Reynolds number, $Re_a>0$, breaks the symmetry of the particle \lb{pair }trajectories~\citep{kulmor_jfm08,pic_etal_prl13} and induces an anisotropic microstructure, in turns responsible of shear-thickening. 

It is well established that the macroscopic flow behaviour  changes dramatically
from the laminar conditions to the typical chaotic dynamics of transitional and turbulent flows
when increasing the Reynolds number, still for single phase fluids. The effect of a dense suspended phase
on the transition to turbulence in pipe flows has been investigated experimentally 
by \cite*{matas2003transition}. These authors report a non-trivial behaviour of
the critical Reynolds number at which transition is observed. The critical Reynolds number for relatively large particles is found
to first decrease and then increases, with a minimum in the range $\Phi\sim0.05-0.1$. This non-monotonic behaviour cannot
be explained only in terms of the increase of the suspension effective viscosity. 
These experiments have been numerically reproduced in \cite{Yu13}. Recently, \cite{lashgari2014flow} showed that the 
flow behaviour is more complex than that pertaining to unladen flows:  three different regimes coexist with different probability when changing the volume fraction $\Phi$ and the Reynolds number $Re$. In each regime the flow is dominated by viscous, turbulent and particle stresses respectively. 

As far as the turbulent regime is concerned, most part of the previous studies pertains to the dilute or very dilute regimes.
In the very dilute regime, the particle concentration is so small that the solid phase has a negligible effect on the
flow. In this, so-called, one-way coupling regime, the main object of most investigations is the particle transport properties. In particular, 
 inertia affects the particle turbulent dispersion leading to preferential migration. 
Small-scale clustering has been observed both in isotropic, 
see e.g.~\citep{toschi2009lagrangian} and inhomogeneous flows, see e.g.~\citep{sarcha}. It amounts to 
a segregation of the particles in fractal sets~\citep{becetal,toschi2009lagrangian} induced by the coupling of the turbulent flow dynamics (dissipative) 
and the particle inertia when the time-scales of the two phenomena are similar. 
In wall bounded flow, particle inertia induces a mean particle drift towards the wall, so-called 
turbophoresis~\citep{reeks}. This effect is most pronounced when the particle inertial time scale almost matches the turbulent near-wall 
characteristic time~\citep{soldati}. Clustering and turbophoresis interact leading to the formation 
of streaky particle patterns~\citep[e.g.][]{sardina2011large}. 

Increasing the solid phase concentration, while still keeping  small the volume fraction and the particle diameter with respect to the flow length scales,  the flow satisfies the so-called two-way coupling approximation, see among others~\cite{ferrante2003physical,balachandar2010turbulent}.
This regime is characterised by high mass density ratios,  i.e.\ the ratio between the mass of the solid  phase and the \lb{fluid } one, and low volume fractions~\citep{balachandar2010turbulent} in the limit of high mass fractions;
this occurs typically for solid particles or droplets dispersed in a gas phase when the density ratio between particles and fluid  is 
high (about 1000). In this regime the dispersed phase back-reacts on the carrier fluid exchanging momentum, being inter-particle
interactions and excluded volume effects negligible given the small volume fractions. 
In homogeneous and isotropic flows, \cite{squires1991preferential} and \cite{elghobashi1993two} 
observe an attenuation of the  turbulent kinetic energy at large scales accompanied by
an energy increase at small scales.  \cite{sundaram1999numerical} and \cite{ferrante2003physical}  also performed systematic studies to understand the effect of the particle
inertia and of the mass fraction on the flow.
\cite{jfm_2way} report that the particle segregation in anisotropic fractal sets induces an alternative mechanism to 
directly transfer energy from large to small scales. 
Similar results have been reported for wall-bounded turbulent flows. \cite*{kulick1994particle} showed that
the solid phase reduces the turbulent near-wall fluctuations increasing their anisotropy, see also \cite{li2001numerical}.
\cite*{zhao2010turbulence} showed how these interactions may lead to drag reduction.

If the dispersed phase is not constituted by elements smaller than the hydrodynamic scales, the 
suspended phase directly affects the turbulent structures at scales similar or below the particle size~\citep{naso2010interaction,bellani712shape,homann2013effect}. 
Being
the system non-linear and chaotic, these large-scale interactions modulate the whole process inducing non-trivial effects on 
the turbulence cascade~\citep*{lucci2010modulation,yeo2010modulation} where increase or decrease of the spectral energy distribution depends on the particle size and mass fraction. \red{\cite{pan1996numerical} were the first to simulate the effect
of finite-size particles in a turbulent channel flow showing that when these are larger than the dissipative scale turbulent
fluctuations and stresses become larger.} 
The open channel flow laden with heavy finite-size particles has been investigated in the dilute regime 
by \cite{kidanemariam2013direct} and \cite{kidanemariam2014direct} showing that the solid phase preferentially accumulates in near-wall low-speed streaks, the flow structures characterised by smaller streamwise velocity. 

Increasing the volume fraction,
the coupling among the phases becomes richer and particle-particle  hydrodynamic interactions 
and collisions cannot be neglected. 
In this dense regime, so-called four-way coupling, the rheological 
properties of the suspension interact with the chaotic dynamics of the fluid phase when the flow inertia is sufficiently large,
i.e.\ at high Reynolds number. Few studies investigate dense suspensions in the highly inertial regime: \cite{matas2003transition,
loisel2013effect,Yu13} show the effect on transition in wall bounded flows showing a decrease of the critical Reynolds number in
the semi-dilute regime. Concerning the turbulent regime of relatively dense suspensions of wall-bounded flows, \cite*{shao2012fully} report
results for channel flow up to $7\%$ volume fraction both considering neutrally buoyant and heavy particles. These authors 
document a decrease of the fluid streamwise velocity fluctuation due to an attenuation of the large-scale streamwise vortices.
In the case of heavy, sedimenting, particles, the bottom wall behaves as a rough boundary with particles free to re-suspend. 
Different regimes have been observed when the importance of the particle buoyancy is varied in the recent study of 
\cite*{vowinckel2014fluid}. 

In this context, even restricting to the case of neutrally-buoyant particles, little is know on the effect of a dense suspended phase
on the fully turbulent regime. The main reason can be ascribed to the well known difficulties to tackle this case 
 either experimentally or numerically.
As previously noticed, the dense regime is characterised by a complex particle
microstructure that induces non-trivial macroscopic  features. When the large-scale inertia is high enough, the 
interaction between the suspension microstructure, i.e.\ rheology, and turbulence dynamics is expected to 
significantly alter the macroscopic flow dynamics. This is the object of the present study. 

To this end, we consider  
turbulent channel flows laden with finite-size particles (radius $a=h/18$ with $h$ the half-channel height) 
up to a volume fraction $\Phi=0.2$. We use data from a Direct Numerical Simulation that fully describe the 
solid phase dynamics via an Immersed Boundary Method.
We show that the classical laws for the turbulent mean velocity
profiles are modified in the presence of the particles and the overall drag increases.  
At the highest volume fraction investigated, $\Phi=0.2$, 
the velocity fluctuation intensities and the Reynolds shear stresses are found to 
suddenly decrease. We consider the mean momentum budget to show that the particle-induced stress is responsible
of the overall drag increase  at high $\Phi$, while the turbulent drag decreases.

\section{Methodology}

\subsection{Numerical Algorithm}

\red{During the last years, different methods have been proposed to perform accurate Direct Numerical Simulations of 
 dense multiphase flows. Fully Eulerian methods have been adopted to deal with two-fluid flows, such as front-tracking, sharp- or diffuse interface methods see e.g.\ 
 \cite{tryggvason2001front,bray2002theory,celani2009phase,benzi2009mesoscopic,magaletti2013sharp}, whereas mixed Lagrangian-Eulerian techniques are found to 
 be the most appropriate for solid-liquid suspensions \citep{ladd2001lattice,takagi2003physalis,lucci2010modulation,vowinckel2014fluid,kidanemariam2014direct}. }  
In this framework, the present simulations have been performed with a numerical code that fully describes the coupling between the solid and fluid phases \citep{breugem_jcp12}.
The Eulerian  fluid phase evolves according to the incompressible Navier-Stokes equations,
\begin{align}
&\boldsymbol{\nabla}\cdot\mathbf{u}_f = 0,\label{eq:ns1}\\
&\frac{\partial\mathbf{u}_f}{\partial t} +\mathbf{u}_f \cdot \boldsymbol{\nabla}\mathbf{u}_f =
-\frac{1}{\rho}\boldsymbol{\nabla}p + \nu \nabla^2\mathbf{u}_f +\mathbf{f}, 
\label{eq:ns2}
\end{align}
where $\mathbf{u}_f$ is the fluid velocity, $\mathbf f$ a generic force field, 
$p$ the pressure, $\nu=\mu/\rho$ the kinematic viscosity of the pure fluid with
$\mu$ the dynamic viscosity and $\rho$ the fluid density (same as particle density in this study).
The solid phase consists of neutrally-buoyant rigid spheres whose 
centroid linear and angular velocities, $\mathbf{u}_p$ and $\boldsymbol{\omega}_p$, 
are governed by the Newton-Euler Lagrangian equations,
\begin{align}
& \rho {V}_p\frac{d\mathbf{u}_p}{dt}= \rho \oint_{\partial {\cal V}_p}\boldsymbol{\tau}\cdot\mathbf{n}\,dS,\label{eq:ne1}\\
& I_p\frac{d\boldsymbol{\omega}_p}{dt}=\rho \oint_{\partial {\cal V}_p}\mathbf{r}\times\boldsymbol{\tau}\cdot\mathbf{n}\,dS,
\label{eq:ne2}
\end{align}
where $a$ is the particle radius and ${ V}_p=4\pi a^3/3$ the particle volume; the fluid stress is $\boldsymbol{\tau}=-p\mathbf{I}+2 \mu\mathbf{E}$ with  $\mathbf{E}=\left(\boldsymbol{\nabla}\mathbf{u}_f+\boldsymbol{\nabla}\mathbf{u}_f^{T}\right)/2$ the {deformation tensor. In eq.~\eqref{eq:ne2}, $I_p=(2/5) \rho {\cal V}_p a^2$ represents the moment of inertia, $\mathbf r$ the distance vector from
the centroid of the sphere and $\mathbf{n}$ the unity vector normal to the particle surface $\partial {\cal V}_p$.
On the particle surfaces, Dirichlet boundary conditions for the fluid phase are enforced as 
$\mathbf{u}_f|_{\partial {\cal V}_p} =\mathbf{u}_p+\boldsymbol{\omega}_p\times\mathbf{r}$.

In the simulations reported in this paper, the coupling between the two phases is obtained by using an Immersed Boundary Method: this amounts to 
adding a force field $\mathbf f$ on the right-hand side of equation~\eqref{eq:ns2} to mimic the actual boundary condition
at the moving particle surface, i.e. $\mathbf{u}_f|_{\partial {\cal V}_p} =\mathbf{u}_p+\boldsymbol{\omega}_p\times\mathbf{r}$. The fluid phase is evolved solving Eq.\eqref{eq:ns1}-\eqref{eq:ns2} in a domain containing all the particles, without the need to adapt the mesh to the current particle position,
using a second order finite difference scheme on a staggered mesh. The time integration is performed by a third order Runge-Kutta
scheme combined with a pressure-correction method on each sub-step. The Lagrangian evolution of 
Eqs.~\eqref{eq:ne1}-\eqref{eq:ne2} is performed using the same Runge-Kutta scheme of the Eulerian solver. The particle surface 
is tracked using $N_L$ Lagrangian points uniformly distributed on the surface of the spheres on which the forces exchanged with the fluid phase are imposed. 
To maintain accuracy, the right-hand side of equations~\eqref{eq:ne1}-\eqref{eq:ne2} are rearranged in terms of the 
IBM force field and take into account the mass of the fictitious fluid phase occupied by the particle volumes
\begin{align}
& \rho {V}_p\frac{d\mathbf{u}_p}{dt}= - \rho \sum_{l=1}^{N_L} \mathbf{F}_l\, \Delta V_l\, + \rho \frac{d}{dt}\left(\int_{V_p} \mathbf{u}_f dV\right),\label{eq:ne2_b}\\
& I_p\frac{d\boldsymbol{\omega}_p}{dt}=-\rho  \sum_{l=1}^{N_L} \mathbf{r_l}\times\mathbf{F}_l\, \Delta V_l\,+
 \rho \frac{d}{dt}\left(\int_{V_p} \mathbf{r}\times \mathbf{u}_f dV\right),
\label{eq:ne2_b}
\end{align}
where $\Delta V_l$ is the volume of the cell around the $l$ Lagrangian point and $\mathbf{r}_l$ the distance from the particle
centre. $\mathbf{F}_l$ is the force acting on the $l$ Lagrangian point on the particle and is related to the Eulerian force 
field $\mathbf f$: $\mathbf{f}(\mathbf{x})=\sum_{l=1}^{N_L} F_l \, \delta_d (\mathbf{x}-X_l)\Delta V_l$. 
The procedure to determine the force field from the boundary conditions at the particle surface follows an iterative 
algorithm  that allows the code to achieve second order global accuracy in space. 
All the details of this implementation are presented in \cite{breugem_jcp12}. 
  
\begin{figure}
 \begin{center}
 \includegraphics[width=.59\linewidth]{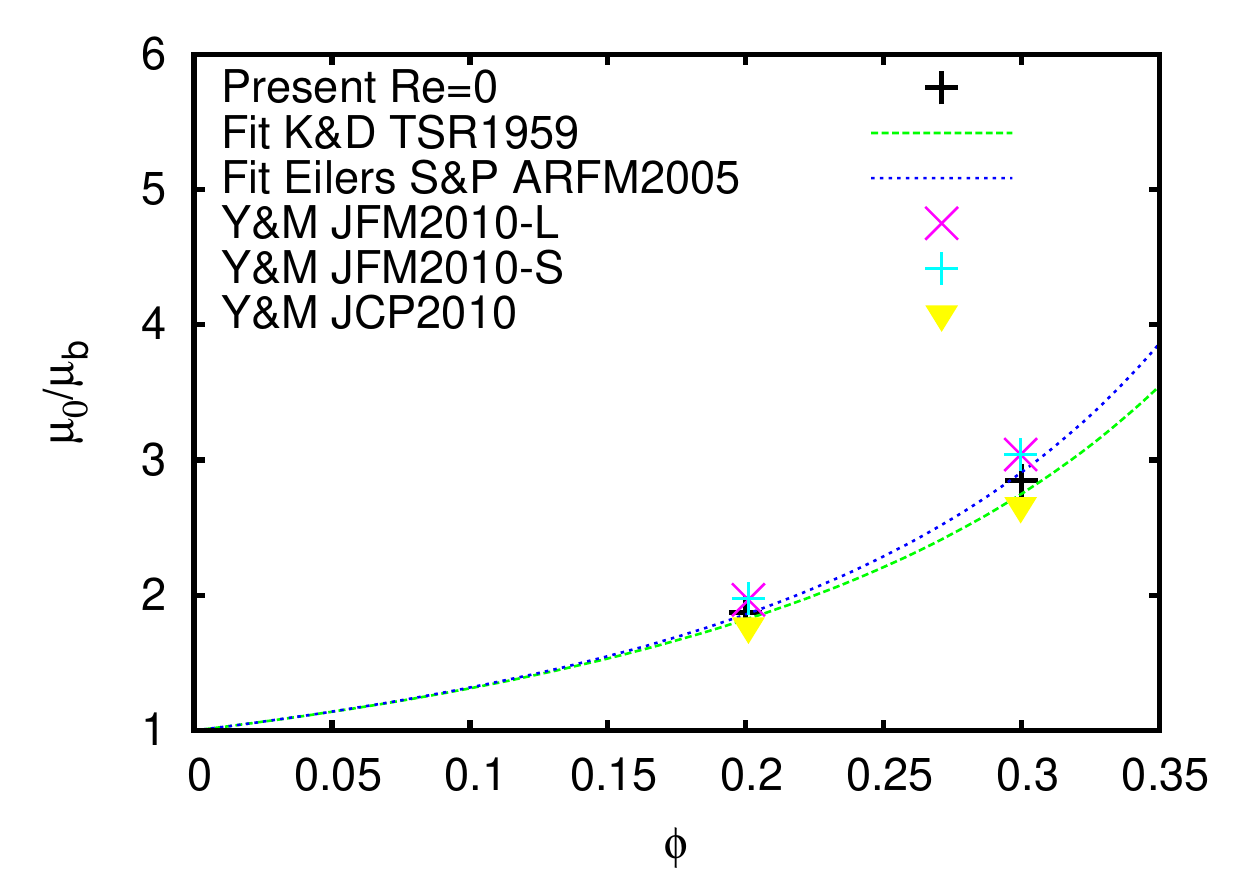} 
 \end{center}
\caption{\label{fig:0} Relative viscosity $\nu_r$ versus the volume fraction $\Phi$ in a Couette flow in absence of inertia, $Re=0$. The present data are  compared with the results by \cite{yeomax_jfm10,yeo2010simulation} and the Eilers Fit  $(1+1.25\Phi/(1-\Phi/0.65))$. }
 \end{figure} 

The numerical method models the interaction among the particles also when their gap distance is of the order or below the grid 
size. In particular, lubrication models based on the Brenner's asymptotic solution~\citep{brenner1961slow} 
are used to correctly reproduce the interaction between particles when their gap distance is smaller than twice the mesh size. 
When particles collide with the wall or among themselves a soft-collision model ensures an almost elastic
rebound with a restitution coefficient set at $0.97$. A complete discussion of these models can be found in \cite{breugem_jcp12} and \cite{lambert2013active} where several test cases are presented as validation.   

To avoid duplication of published material, we provide here only evidence for the ability of the present numerical tool to accurately
simulate dense suspensions. Figure~\ref{fig:0} displays the relative viscosity, 
the ratio between the effective viscosity of the suspension and the viscosity of the 
fluid phase $\nu_r=\nu_e/\nu$,  in laminar flows for two different volume fractions, $\Phi=0.2$ and $\Phi=0.3$.  The 
configuration where this is measured is the Couette flow at vanishing Reynolds number where the wall-to-wall distance is ten times  the particle radius. A cubic mesh is used to discretise the computational domain with 8 point per particle radius, $a$. The  streamwise and spanwise length of the computational domain are 1.6 times the wall-normal width, i.e.\ 16$a$. The relative viscosity extracted after the  initial transient phase is measured by the friction at the wall and perfectly matches previous numerical investigations 
\citep{yeomax_jfm10,yeo2010simulation} and empirical fits of experimental data, like the Eilers Fit \citep{stipow_arfm05}.

\subsection{Flow configuration}


\begin{table}
\centering
\begin{tabular}{c|cccc}
$\Phi$&0.0&0.05&0.1&0.2\\
\hline
$N_p$&0&2.500&5.000&10.000\\
\hline
$L_x\times L_y \times L_z$&
\multicolumn{4}{c}{$6h\times2h\times3h$}\\
\hline
$N_x\times N_y \times N_z$ &
\multicolumn{4}{c}{$864\times288\times432$}\\
\hline
$Re_b$&
\multicolumn{4}{c}{5600}\\
\hline
$\nu_r$&1.0&1.14&1.33&1.89\\
\hline
$Re_e$&5600&4912&4210&2962
\\
\hline
\end{tabular}
\caption{Summary of the Direct Numerical Simulations reported here. They pertain to suspensions of $N_p$ particles of
radius $a/h=1/18$ at different volume fractions $\Phi$. 
$N_x, N_y, N_z$ indicate the number of grid points in each direction and the bulk Reynolds number is defined as $Re_b=U_0*2h/\nu$.
The relative viscosity, i.e. the ratio between 
effective suspension viscosity and the fluid viscosity $\nu_r=\nu_e/\nu=[1+ 1.25*\Phi/(1-\Phi/0.6)]^2$ has been estimated
via the Eilers fit~\citep{stipow_arfm05}. The effective bulk Reynolds number is defined as $Re_e=U_0 2h/\nu_e=U_0 2h/(\nu \,\nu_r)=
Re_b/ \nu_r$.  
\label{tab:0}}
\end{table}

In this work we study a pressure driven channel flow between two infinite flat walls 
located at $y=0$ and $y=2h$ with $y$ the wall-normal direction. 
Periodic boundary conditions are imposed in the  streamwise, $x$, and spanwise, $z$, directions for a domain size of $L_x=6h$, $L_y=2h$ and $L_z=3h$. A mean pressure gradient acting in the streamwise direction imposes a fixed value of the bulk velocity $U_0$ across the channel corresponding to a constant bulk Reynolds number 
 $Re_b=U_0 2h/\nu=5600$, with $\nu$ the kinematic viscosity of the fluid phase; this value corresponds to a Reynolds number based on the friction velocity $Re_{\tau}= U_* h/\nu=180$ for the single phase case where $U_*=\sqrt{\tau_w/\rho}$ with $\tau_w$ the 
 stress at the wall. As reported in table~\ref{tab:0}, the bulk Reynolds number based on the suspension 
 effective viscosity $Re_e$ varies with the volume fraction following the increase of the effective viscosity $\nu_e=\nu_r\,\nu$ where $\nu_r$ is the 
 relative viscosity estimated by the Eilers fit~\citep{stipow_arfm05}. 
The domain is discretised by a cubic mesh of $864\times288\times432$ points in the streamwise, wall-normal and spanwise directions.  Hereafter all the variables 
have been made dimensionless 
with $U_0$ and $h$, except those 
with the superscript ``$^+$''  that are scaled with $U_*$ and $\delta_*=\nu/U_*$ (inner scaling).

Non-Brownian spherical neutrally-buoyant rigid particles are considered. The ratio between the particle 
radius and the channel half-width is fixed 
to $a/h=1/18$, corresponding to 10 plus units for the lowest volume fraction considered and 12 for the largest. 
Three different volume fractions, $\Phi=0.05;0.1;0.2$, have been examined in addition to the single phase case for a direct comparison. The highest volume fraction here addressed requires 10000 finite-size particles in the computational domain
with $N_l=746$ Lagrangian control points on the surface of each sphere and 8 Eulerian grid points per particle radius, see table~\ref{tab:0}.    
The simulations were run on a Cray XE6 system using 2048 cores for a total of about $10^6$ CPU hours for each case.

\begin{figure}
 \begin{center}
 \includegraphics[width=.49\linewidth]{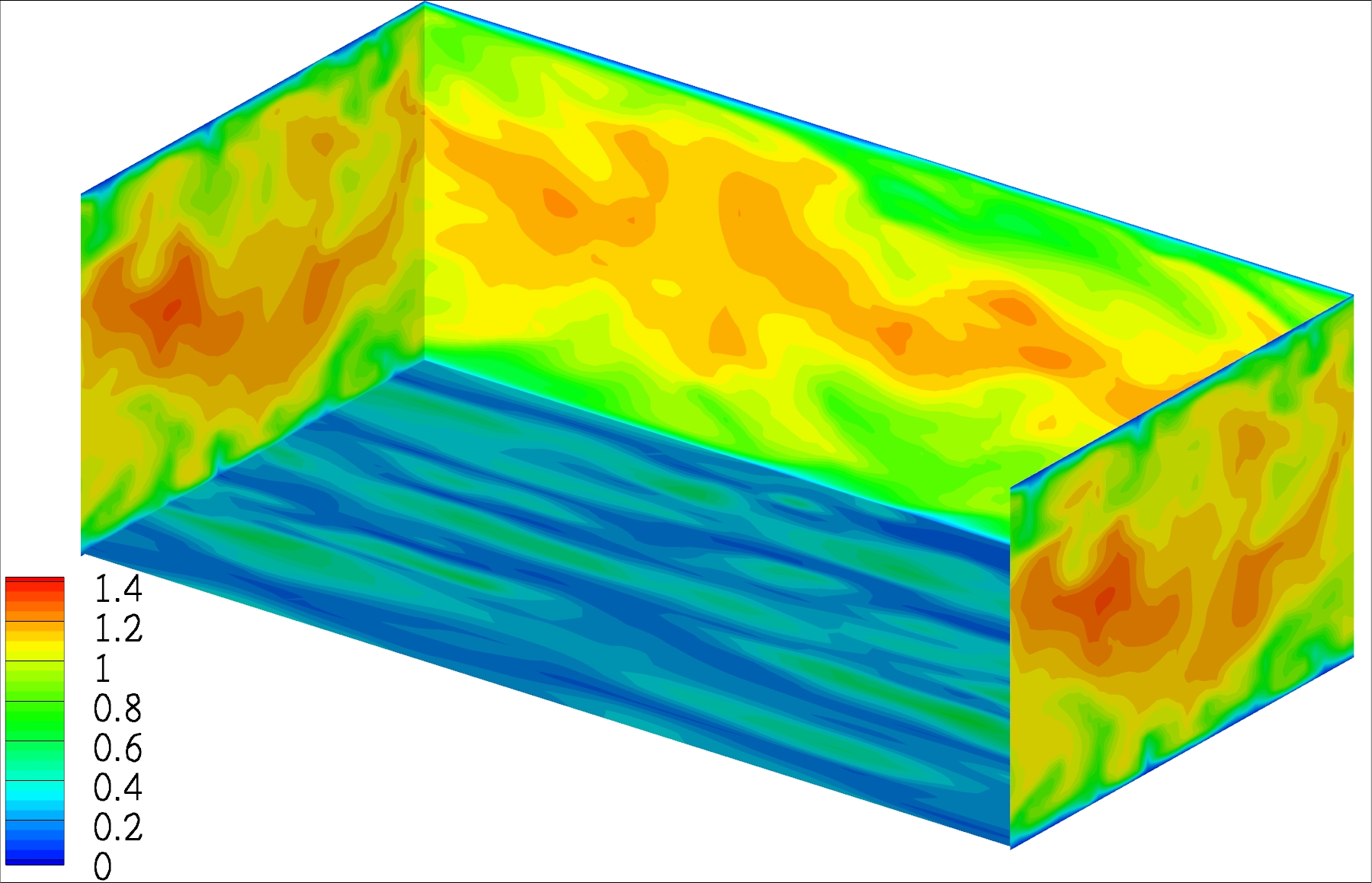} 
\put(-160,10){$\Phi=0$}\hfill
 \includegraphics[width=.49\linewidth]{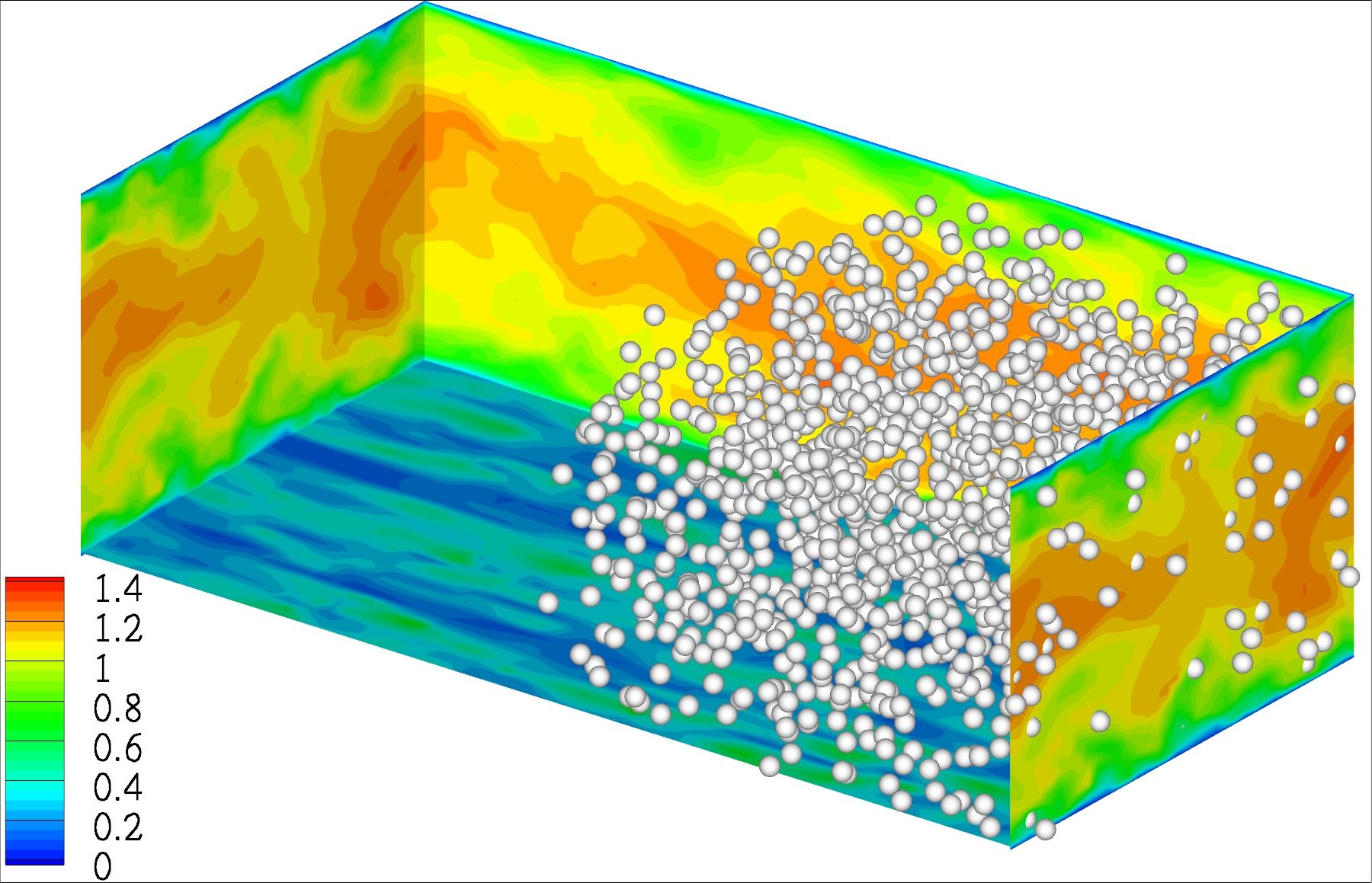}
\put(-160,10){$\Phi=0.05$}\\[.1cm]
 \includegraphics[width=.49\linewidth]{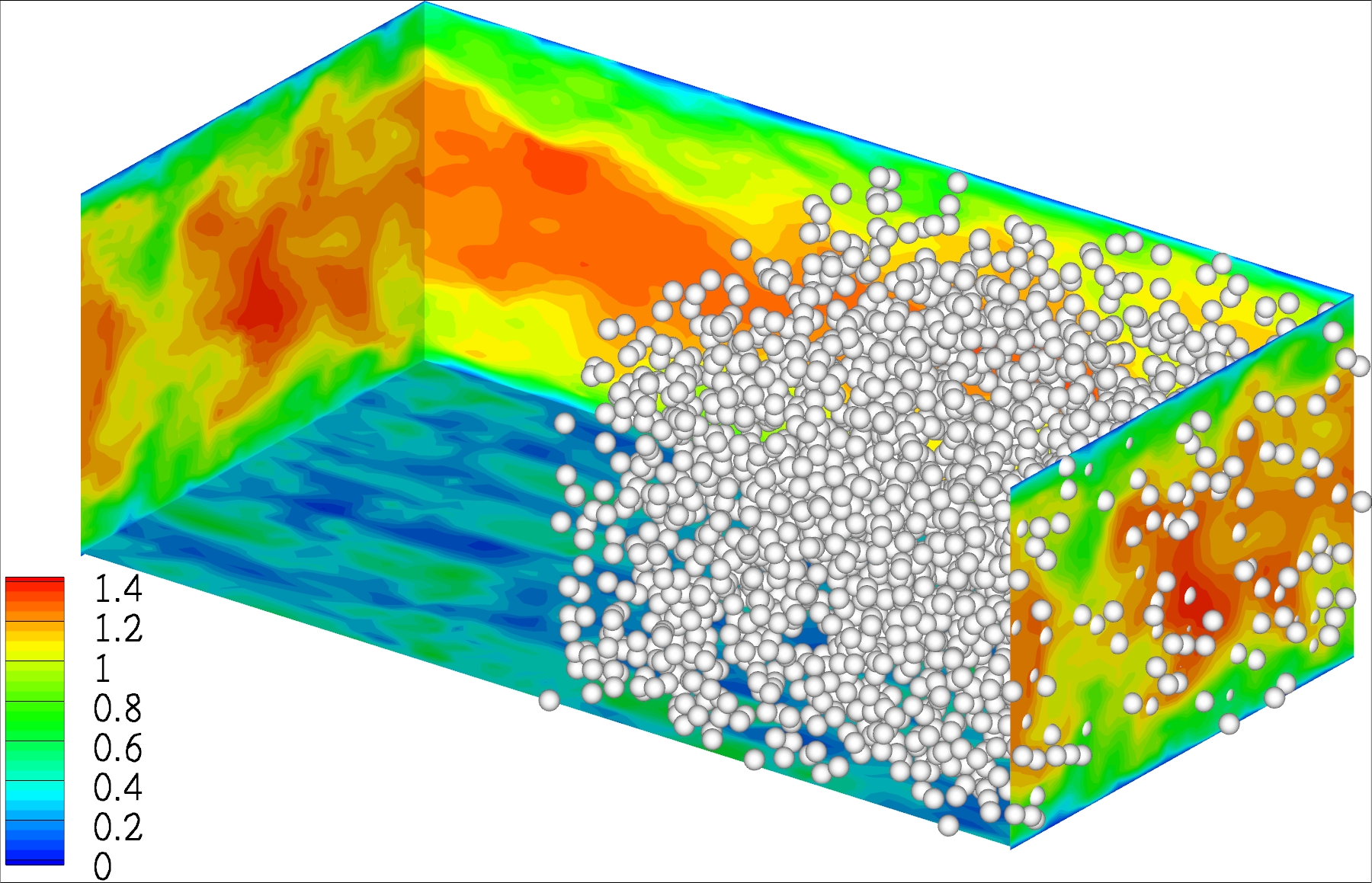}
\put(-160,10){$\Phi=0.1$}\hfill
 \includegraphics[width=.49\linewidth]{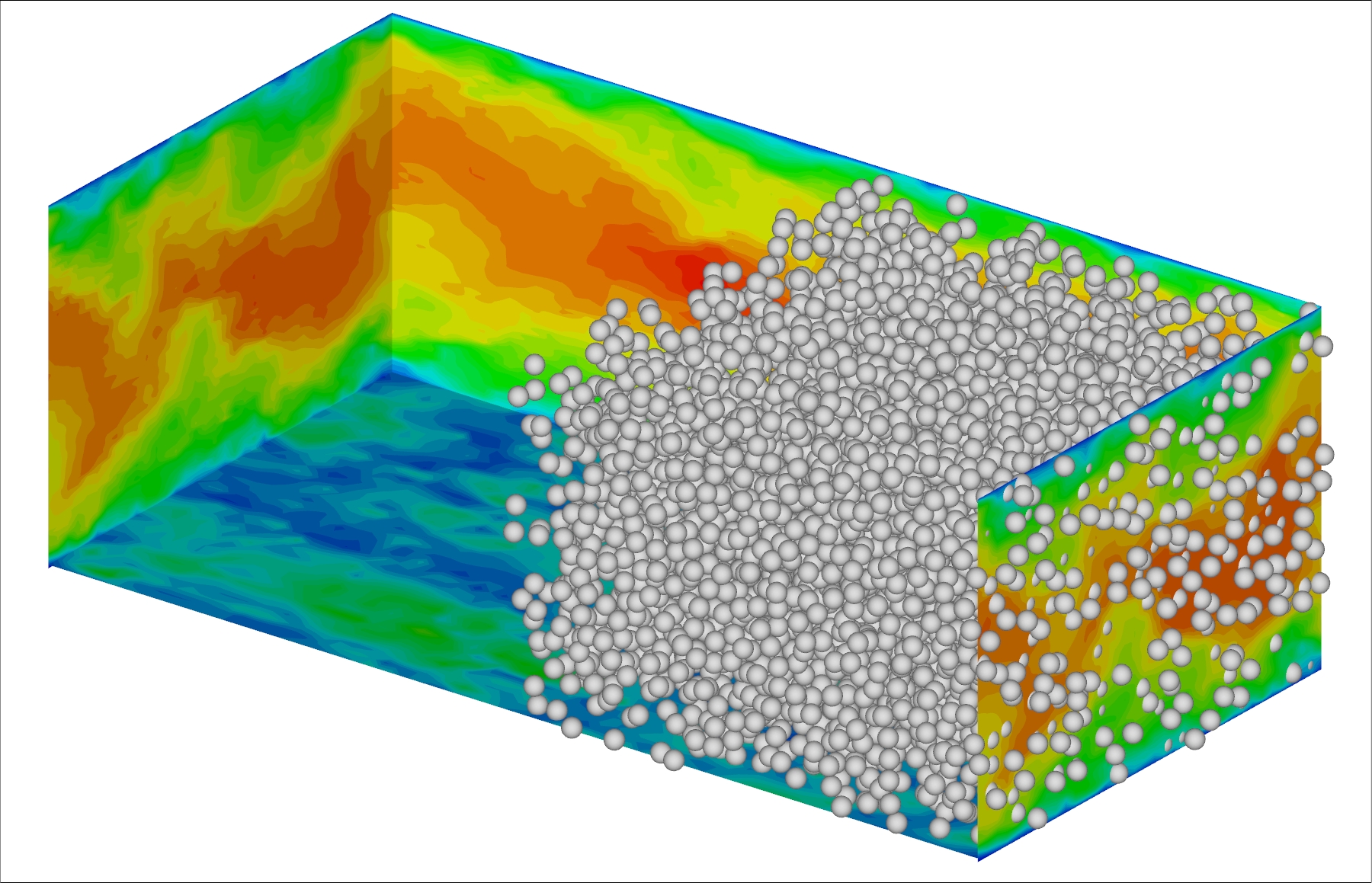}
\put(-160,10){$\Phi=0.2$}
\caption{\label{fig:1} Instantaneous snapshots of the streamwise velocity on different orthogonal planes together with the corresponding 
particle position represented only on one half of the domain. The four panels represent the different values of the volume fraction under investigation,  $\Phi$=0, 0.05, 0.1, 0.2.  }
 \end{center}
 \end{figure}

The simulation starts from the laminar Poiseuille flow for the fluid phase and a random positioning of the particles. Transition
naturally occurs at the fixed Reynolds number because of the noise added by the presence of the particles. Statistics are collected after the initial transient phase.
 %


%
\section{Results}

 \subsection{Single-point flow and particle velocity statistics}


Snapshots of the suspension flow are shown in Figure~\ref{fig:1} for the different \lb{nominal} volume fractions $\Phi$ under investigation. 
The instantaneous streamwise velocity is 
represented on different orthogonal planes with the bottom plane located in the viscous sublayer to highlight the
 low- and high-speed streaks characteristic of near wall turbulence.
Finite-size particles are displayed only on one half of the domain to give a visual feeling on how dense the solid 
 phase is for the different $\Phi$.  Indeed, at the highest volume fraction, $\Phi=0.2$, 
 the particles are so dense that completely hide the bottom wall. 
The cases with $\Phi=0.05$ and $\Phi=0.1$ show velocity contours similar to those of the unladen case where it is possible to recognise the typical near-wall streamwise velocity streaks; these are however more noisy and characterised by significant small scale modulations (of particle size). At $\Phi=0.2$ the small-scale noise is stronger and the streaks become wider.

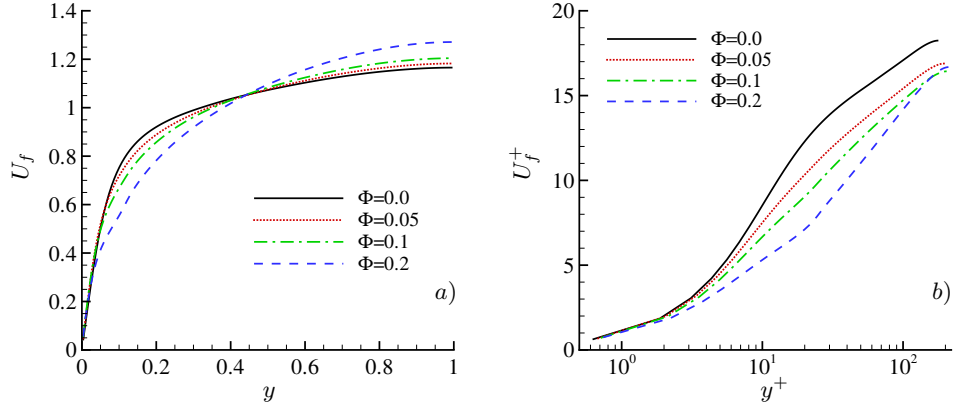
\begin{figure}
 \begin{center}
 \includegraphics[width=.49\linewidth]{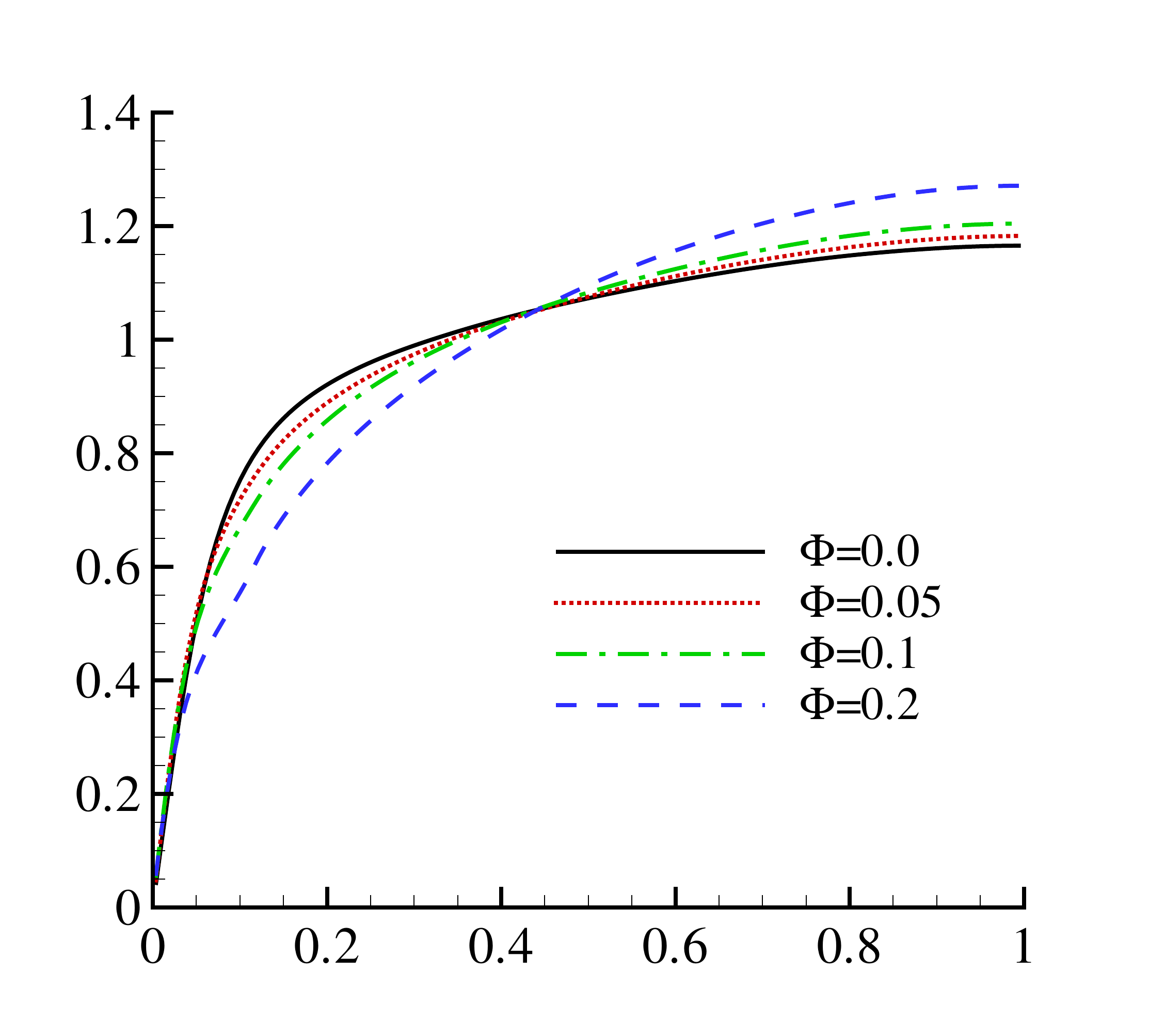} 
\put(-30,40){$a)$}\put(-95,3){$y$}\put(-190,85){\rotatebox{90}{$U_f$}}
 \includegraphics[width=.49\linewidth]{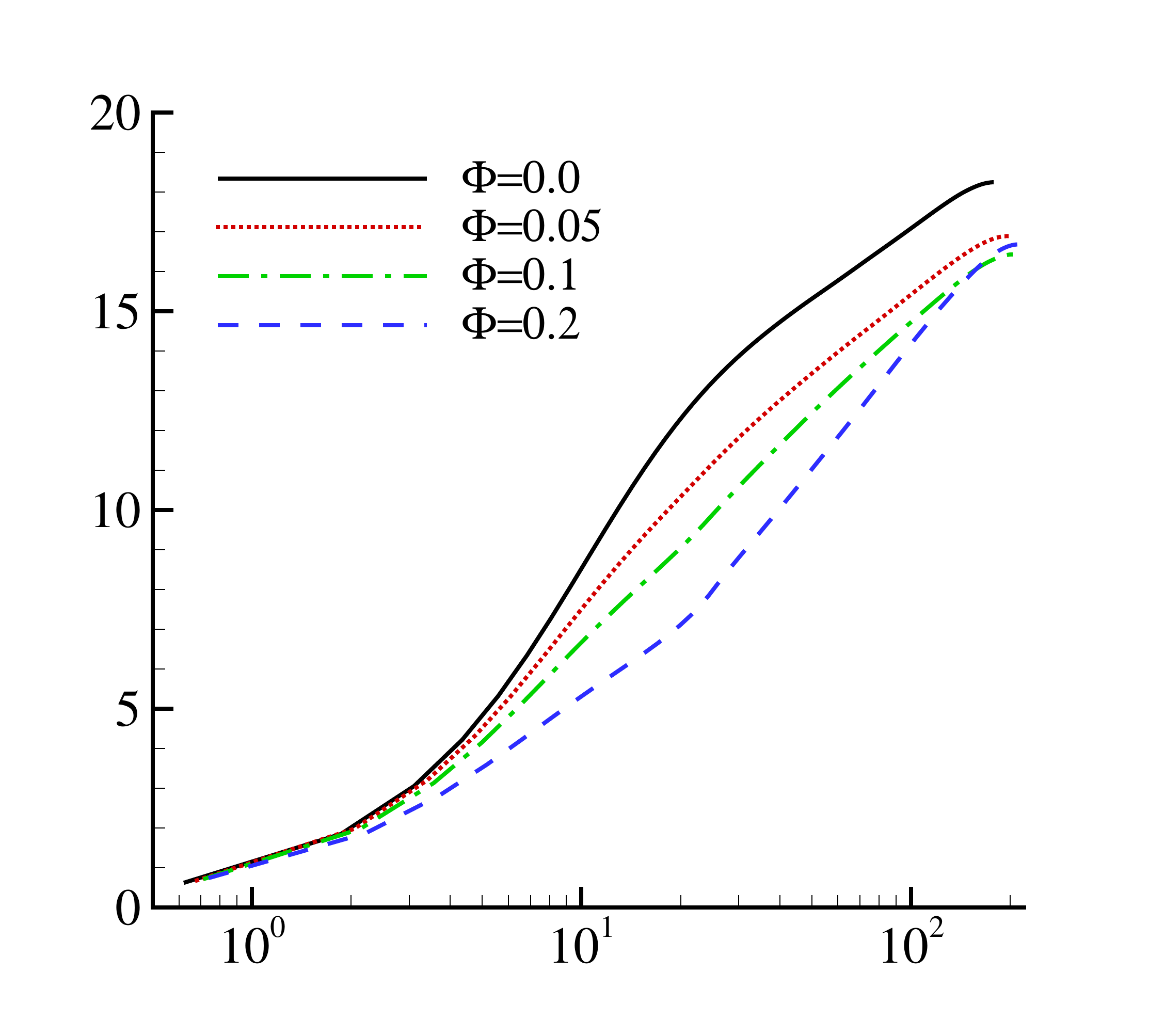} 
\put(-30,40){$b)$}\put(-95,3){$y^+$}\put(-190,85){\rotatebox{90}{$U_f^+$}}
\caption{\label{fig:2} Mean fluid velocity profiles for the different volume
fractions  under investigations in $a)$ outer units \red{(hereafter wall normal distances without the superscript $^+$ are 
assumed to be rescaled by $h$)} and $b)$ inner units: $U_f^+=U_f/U_*$ vs $y^+=y/\delta_*$, with $U_*$ and 
$\delta_*$ the friction velocity and viscous length scale, see definition in the text.}
 \end{center}
 \end{figure}

The mean fluid velocity profiles are shown in figure~\ref{fig:2}. \red{The statistics conditioned to the fluid phase have been
calculated considering only the points located out of the volume occupied by the particles in each field (phase-ensemble
average).}
Panel $a)$ reports the velocity in outer units $U_f$, indicating that the maximum velocity at the mid-plane grows with $\Phi$ (note that the flow rate is constant in these simulations).
In general, the mean velocity more closely resembles the laminar parabolic profile when increasing the volume fraction: the velocity increases in the centre of the channel at higher $\Phi$,
whereas it decreases near the wall, up to $y\sim 1/2$. The higher the volume fraction the more intense this effect is. 
Figure~\ref{fig:2}$b)$ displays the mean fluid velocity profiles scaled in inner units in the log-lin scale, $U_f^+=U_f/U_*$ vs $y^+=y/\delta_*$ where the friction velocity and viscous length are $U_*=\sqrt{\tau_w/\rho}$ and $\delta_*=\nu/U_*$ with 
$\tau_w$ the wall stress. 
The progressive decrease of the profiles
with the volume fraction $\Phi$ indicates that the overall drag increases. Analysing the flow in terms of the canonical classification of wall turbulence, we can still recognise for all 
cases a region ($y^+>40-50$) where the mean profile follows a log-law:
\begin{equation}
U^+=(1/k)\log(y^+)+B
\label{eq:log}
\end{equation}  
with $k$ the von K{\'a}rm{\'a}n constant and $B$ the additive coefficient. Fits of these constants and the corresponding 
Friction Reynolds number $Re_\tau=U_* h /\nu$ are reported
in table~\ref{tab:1} for all the volume fractions investigated. 

The friction Reynolds number computed from the simulation data differs from what can 
be estimated using the rheological properties of the suspension, that is using the relative viscosity $\nu_r$,
see table~\ref{tab:0}. The values of $Re^e_\tau=U_* h/\nu_e=Re_\tau/ \nu_r$ in table~\ref{tab:1}  are computed using the measured wall friction and the effective viscosity of the suspension.
Considering the bulk effective Reynolds number $Re_e=Re_b/\nu_r$, computed in a similar way, it is also possible
to estimate an expected value of the friction Reynolds number using the correlation valid in Newtonian flows, $Re'^e_\tau\simeq0.09 Re_e^{0.88}$ \cite[see][]{pope2000turbulent}.
The data in table~\ref{tab:1} clearly indicate that the effective friction Reynolds number $Re^e_\tau=Re_\tau/\nu_r$ 
is always higher than what expected
considering only the effective viscosity of the suspension, i.e. $Re'^e_\tau$. This fact ($Re'^e_\tau < Re^e_\tau$) implies that the particles alter the turbulence and induce an additional dissipation mechanism, as shown by the higher measured wall friction. 
As shown later, the increased friction can be explained by an increase of the turbulent activity for $\Phi \le 0.1$, whereas this is no more the case for the highest volume fraction considered.
The increased dissipation \lb{at this higher $\Phi$} may be explained by an increased particle induced stress, 
i.e.\ inertial shear-thickening \citep{mor_ra09,pic_etal_prl13}. 
\red{Inertial shear thickening occurs in a dense suspension when inertial effects are present at the particle scale (finite
particle Reynolds number) and amounts to an increase of the effective viscosity with respect the value obtained by rheological
  experiments at vanishing inertia (Reynolds number) and same volume fraction. 
The relatively high Reynolds number of the present turbulent cases triggers inertial effects in the transported particles. 
}


The slope of the log-layer increases,
i.e.\  the von K{\'a}rm{\'a}n constant $k$ decreases, while the additive constant $B$ decreases. At $\Phi=0.2$ the differences with
respect to the unladen case become critical with $B$ strongly negative and $k$ about half of the value for the single phase flow.
These two behaviours act in opposite way: a reduced von K{\'a}rm{\'a}n constant $k$ usually denotes drag 
reduction~\citep{virk}, while a small or negative additive constant $B$ an increase of the drag. The combination of these two counteracting 
effects lead to an increase of the overall drag for the present cases as demonstrated by the increase of the friction Reynolds number 
$Re_\tau$. 
The decrease of the additive constant $B$ appears to be linked to particle-fluid interactions occurring near the
wall. In particular, focusing on the case at $\Phi=0.2$, we note a sudden change in the mean velocity profile
after the first layer of particles, i.e.\ $y^+\sim20\sim d_p^+$. The near wall dynamics
is therefore influenced by the particle layering induced by the wall. A similar behaviour has been observed in turbulent
flows over porous media~\citep{breugem2006influence} suggesting that the {near-wall} layers of particles may act as a porous 
media for the fluid phase.

\begin{table}
\centering
\begin{tabular}{c|cccc}
$\Phi$&0.0&0.05&0.1&0.2\\
\hline
$Re_\tau$&180&195&204&216\\
\hline
$k$&0.4&0.36&0.32&0.22\\
\hline
$B$&5.5&2.7&0.27&-6.3\\
\hline
$Re^e_\tau$&180&171&153&114\\
\hline
$Re'^e_\tau$&180&159&139&102\\
\end{tabular}
\caption{The Von K{\'a}rm{\'a}n constant $k$ and additive constant $B$ of the log-law estimated from the present simulations for the different volume
fractions $\Phi$ examined. $B$ and $k$ have been fitted in the range $y^+\in[50,150]$.
The friction Reynolds number $Re_\tau=U_* h/\nu$,
 and the effective friction Reynolds number, defined as $Re^e_\tau=U_* h/\nu_e
=Re_\tau/ \nu_r$, are also reported together with an estimate of the effective friction Reynolds number based on the correlation $Re'^e_\tau\simeq0.09 Re_e^{0.88}$, see e.g.~\cite{pope2000turbulent}.
\label{tab:1}}
\end{table}

It is worth commenting at this point that increasing the bulk Reynolds number usually leads to a widening of the log-law region and, consequently,  to a stronger impact of the slope of the log-law on the overall mean velocity profile. Assuming that
the constant $B$ does not change significantly upon increasing the bulk Reynolds number (at fixed $d^+$), the
overall mass flux may increase leading to drag reduction if the log region is long enough for the mean velocity at $\Phi=0.2$ to become larger than the corresponding values for the single phase fluid near the channel centreline. This is just a speculation and its proof is out of the scope of the present investigation where we 
consider only a fixed bulk Reynolds number. Simulations at higher Reynolds number and fixed particle size (in plus units) are currently computationally too expensive and out of our reach. 

 \begin{figure}
 \begin{center}
 \includegraphics[width=.49\linewidth]{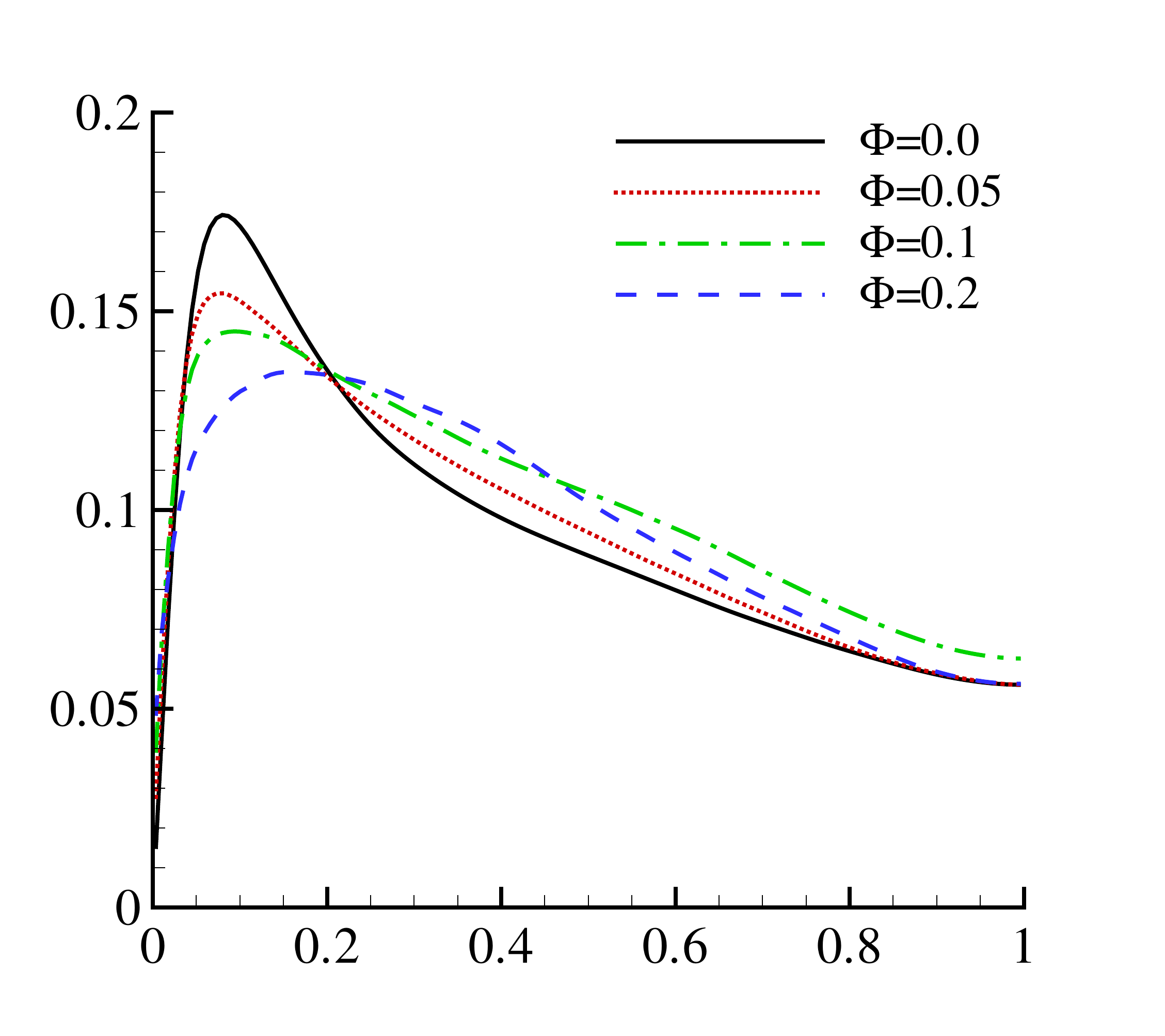} 
\put(-30,40){$a)$}\put(-95,3){$y$}\put(-198,80){\rotatebox{90}{${u_f'}_{rms}$}}\hfill
 \includegraphics[width=.49\linewidth]{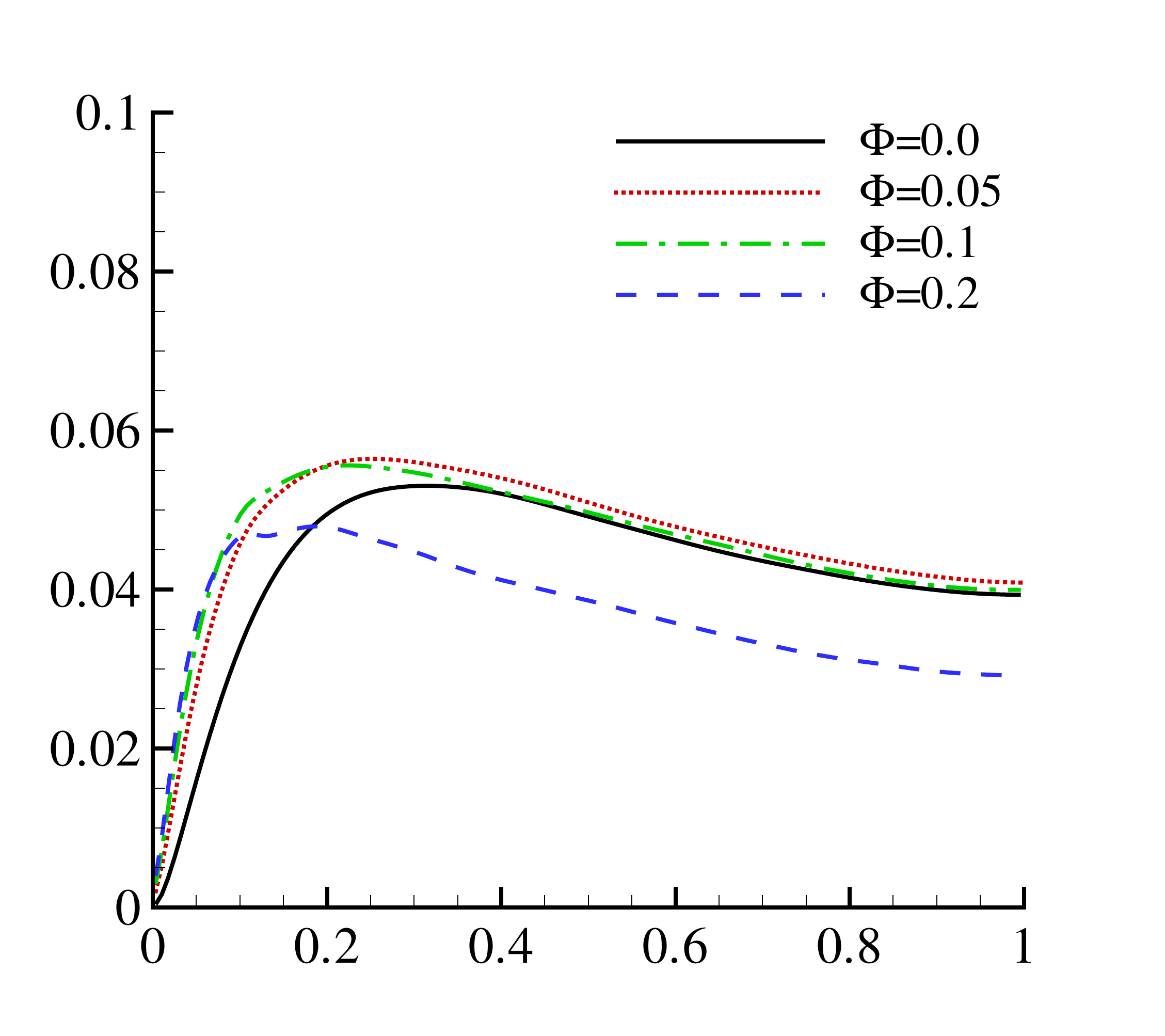} 
\put(-30,40){$b)$}\put(-95,3){$y$}\put(-198,80){\rotatebox{90}{${v_f'}_{rms}$}}\\[.1cm]
 \includegraphics[width=.49\linewidth]{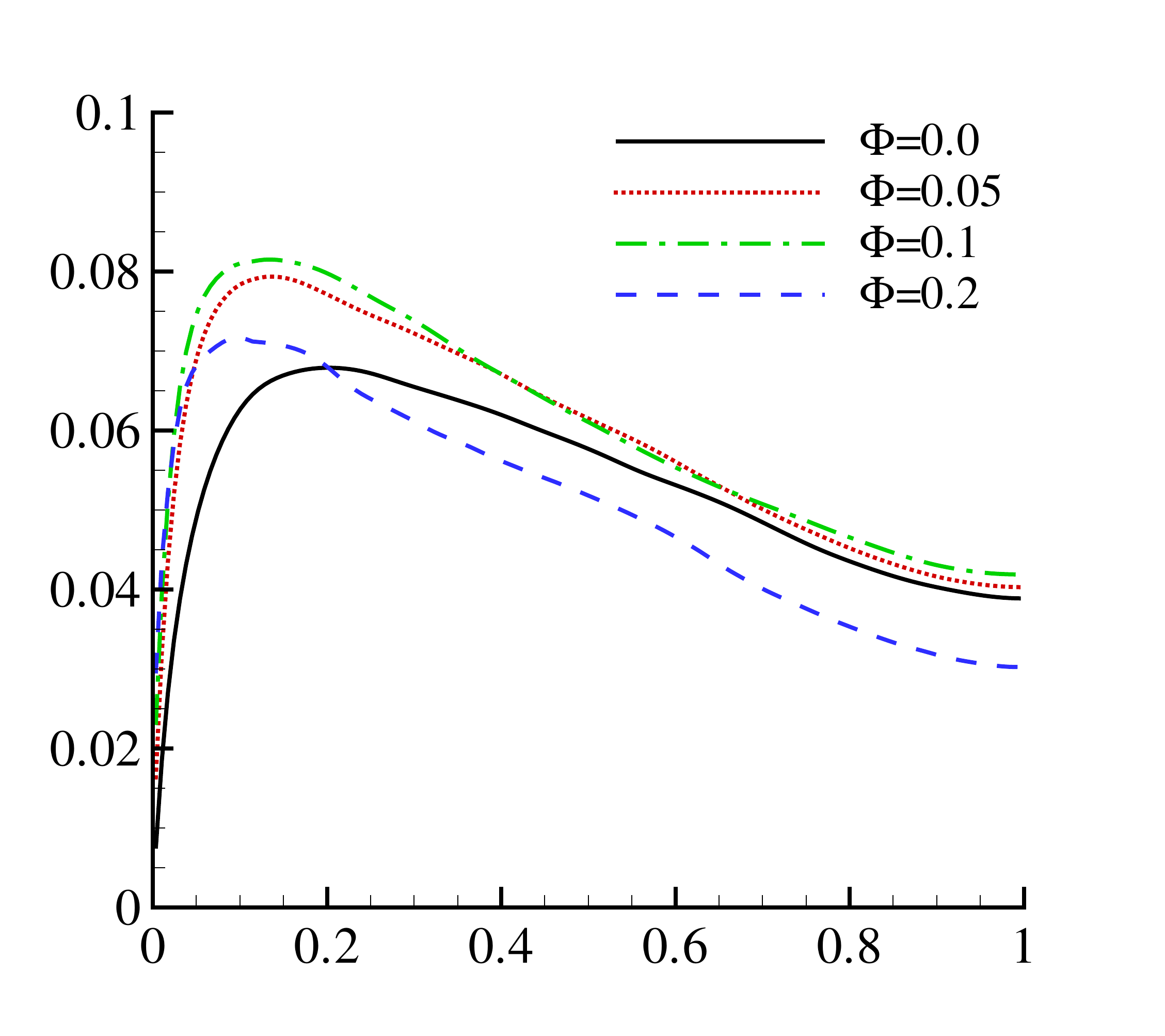} 
\put(-30,40){$c)$}\put(-95,3){$y$}\put(-198,80){\rotatebox{90}{${w_f'}_{rms}$}}\hfill
 \includegraphics[width=.49\linewidth]{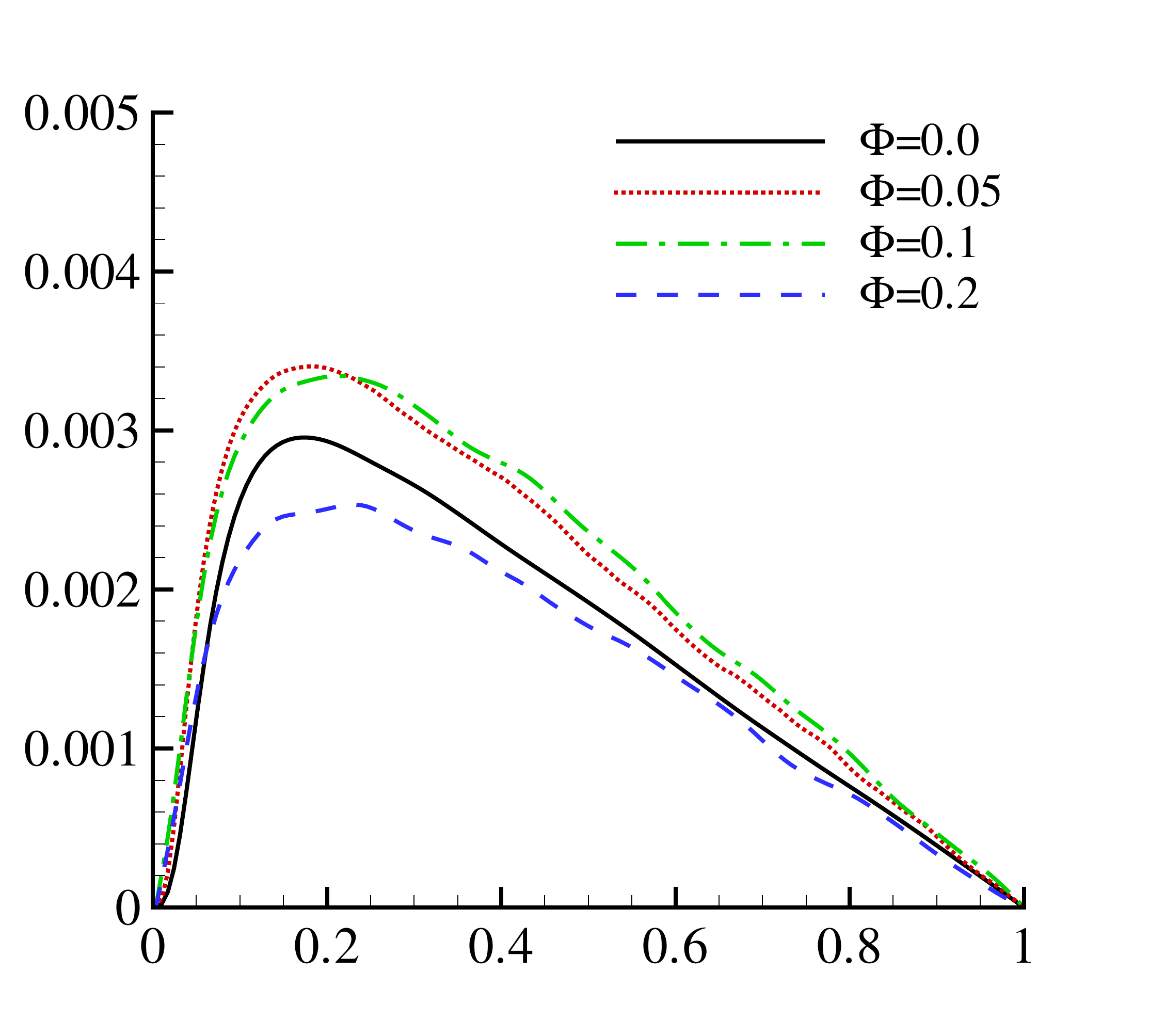} 
\put(-30,40){$d)$}\put(-95,3){$y$}\put(-198,78){\rotatebox{90}{$\langle u_f' v_f'\rangle$}}
\caption{\label{fig:3} Intensity of the fluctuation velocity components
and the Reynolds shear stress for the fluid phase in outer units for different volume fraction
$\Phi$. Panel $a)$ streamwise ${u_f'}_{rms}$; $b)$ wall-normal ${v_f'}_{rms}$;
 $c)$ spanwise ${w_f'}_{rms}$ velocity fluctuations;
  $d)$ shear-stress $\langle u_f' v_f'\rangle $.
}
 \end{center}
 \end{figure}
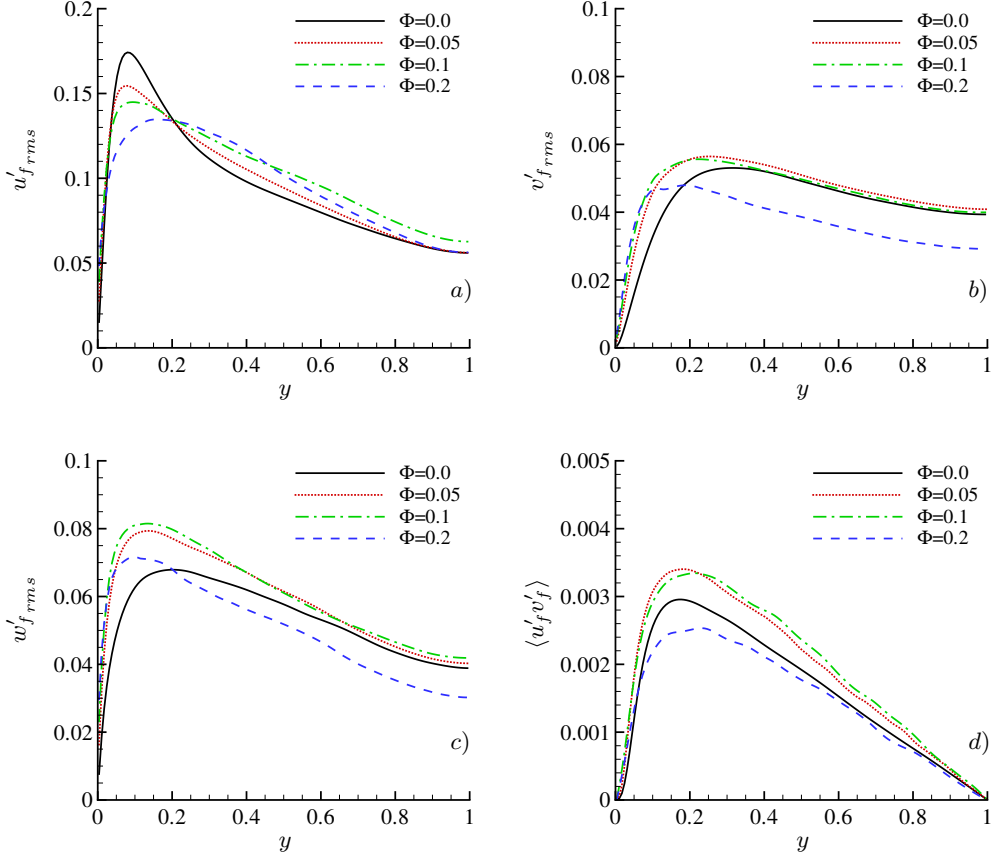 

The root-mean-square (rms) of the fluid velocity fluctuations and the Reynolds shear
stress in outer units are reported in figure~\ref{fig:3}. We note that despite the increase of the friction Reynolds number the peak of the
streamwise velocity rms, ${u'_f}_{rms}$, decreases with $\Phi$, while a non monotonic behaviour is apparent in the
bulk of the flow. For values of $\Phi \le 0.1$ the intensity of the cross-stream velocity fluctuations increases with respect to the single phase cases, 
displaying also higher peak values. This indicates that the particle presence redistributes energy towards a more isotropic state. Interestingly, at the highest volume fraction considered, $\Phi=0.2$, we note a decrease of the level of fluctuations with respect to all the other cases, with the exception of a thin region
close to wall, which will be discussed more in detail in the following. At this high volume fraction we therefore note a reduced
turbulence activity, as confirmed by considering the variations of the Reynolds stress in the presence of particles in panel $d)$ of the same figure. Note that the Reynolds stresses represent the main engine for the production
of turbulent fluctuations. While these stresses increase for $\Phi=0.05$ and $\Phi=0.1$, they decrease at $\Phi=0.2$ despite the increase
of the friction Reynolds number. {At first sight, this aspect may appear controversial, however, as we will discuss
in detail in \S~\ref{sec:bud},  the reduction of the turbulent activity at $\Phi=0.2$ is associated with an increase of the stresses
induced by the solid phase which results in enhanced drag.}   

 \begin{figure}
 \begin{center}
 \includegraphics[width=.49\linewidth]{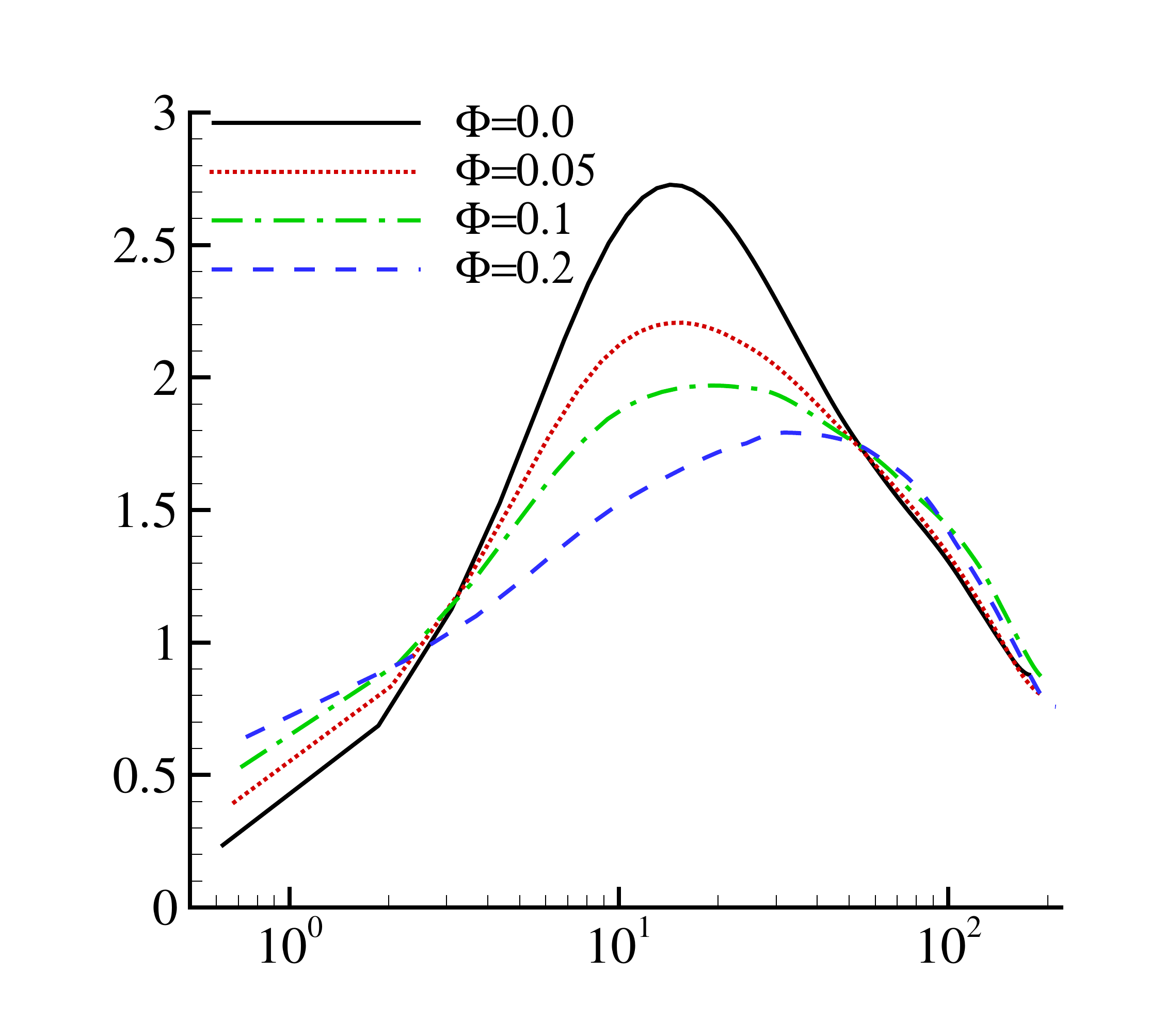} 
\put(-30,40){$a)$}\put(-90,3){$y^+$}\put(-185,80){\rotatebox{90}{${u_f'}^+_{rms}$}}\hfill
 \includegraphics[width=.49\linewidth]{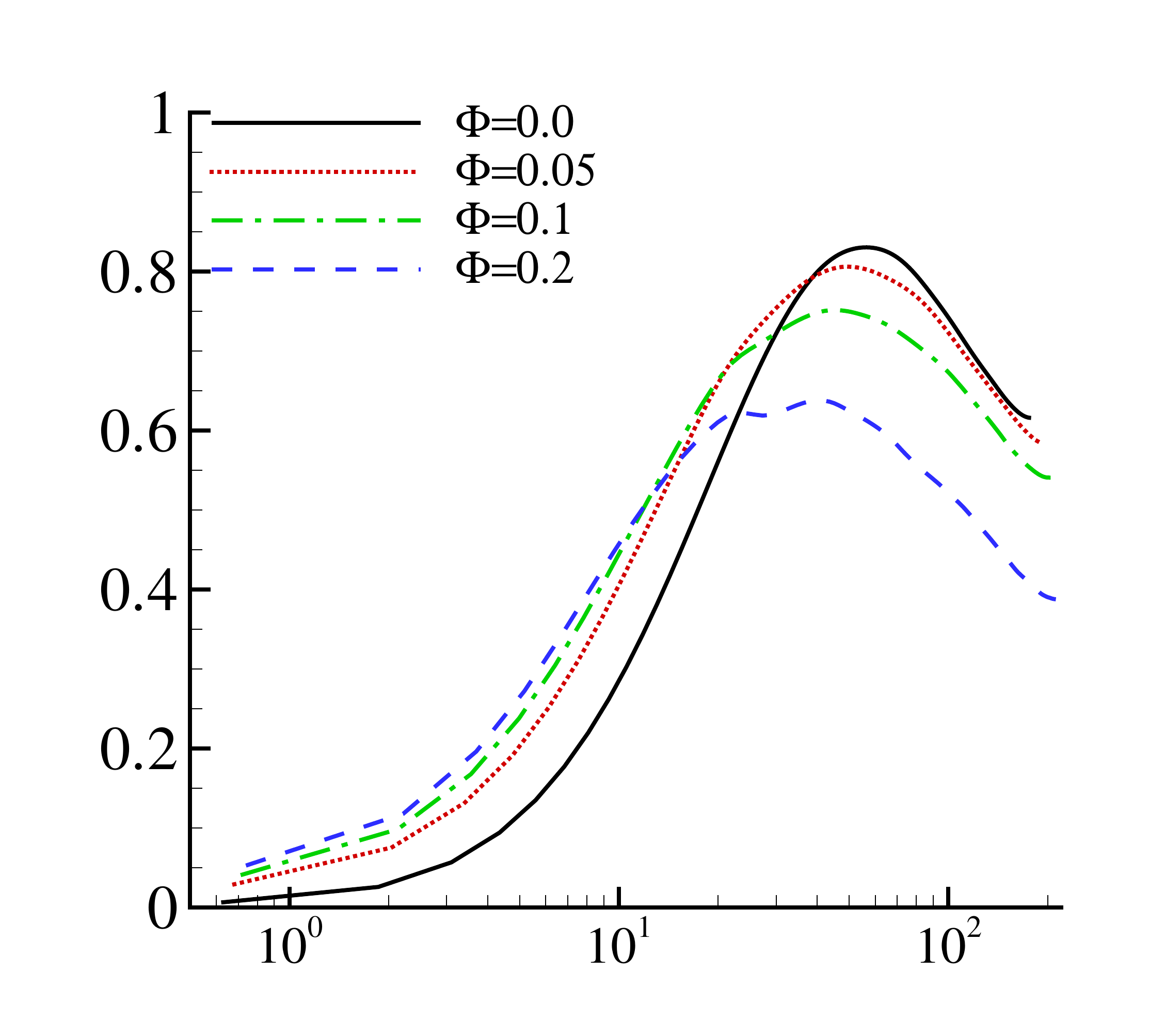} 
\put(-30,40){$b)$}\put(-90,3){$y^+$}\put(-185,80){\rotatebox{90}{${v_f'}^+_{rms}$}}\\[.1cm]
 \includegraphics[width=.49\linewidth]{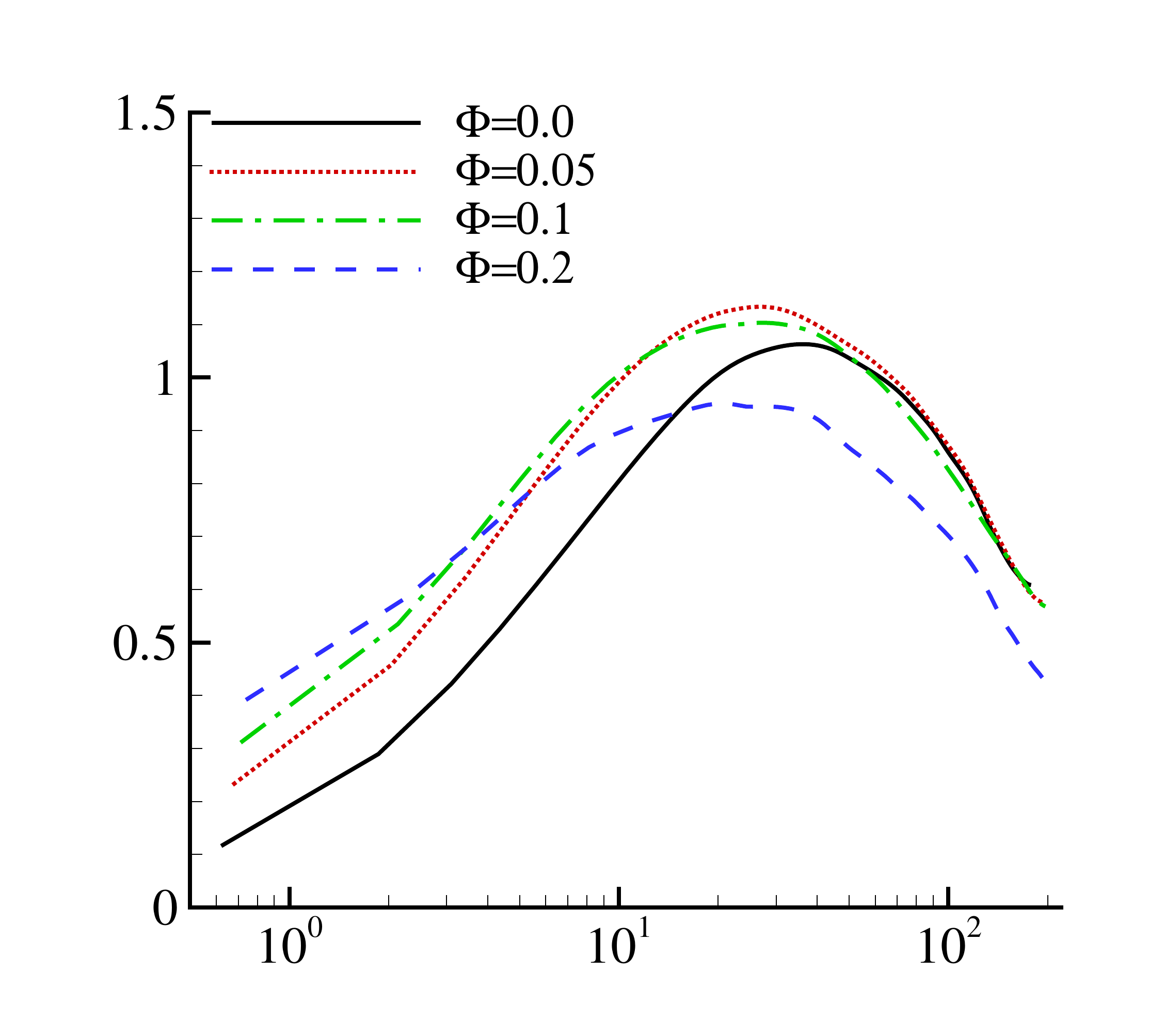} 
\put(-30,40){$c)$}\put(-90,3){$y^+$}\put(-185,80){\rotatebox{90}{${w_f'}^+_{rms}$}}\hfill
 \includegraphics[width=.49\linewidth]{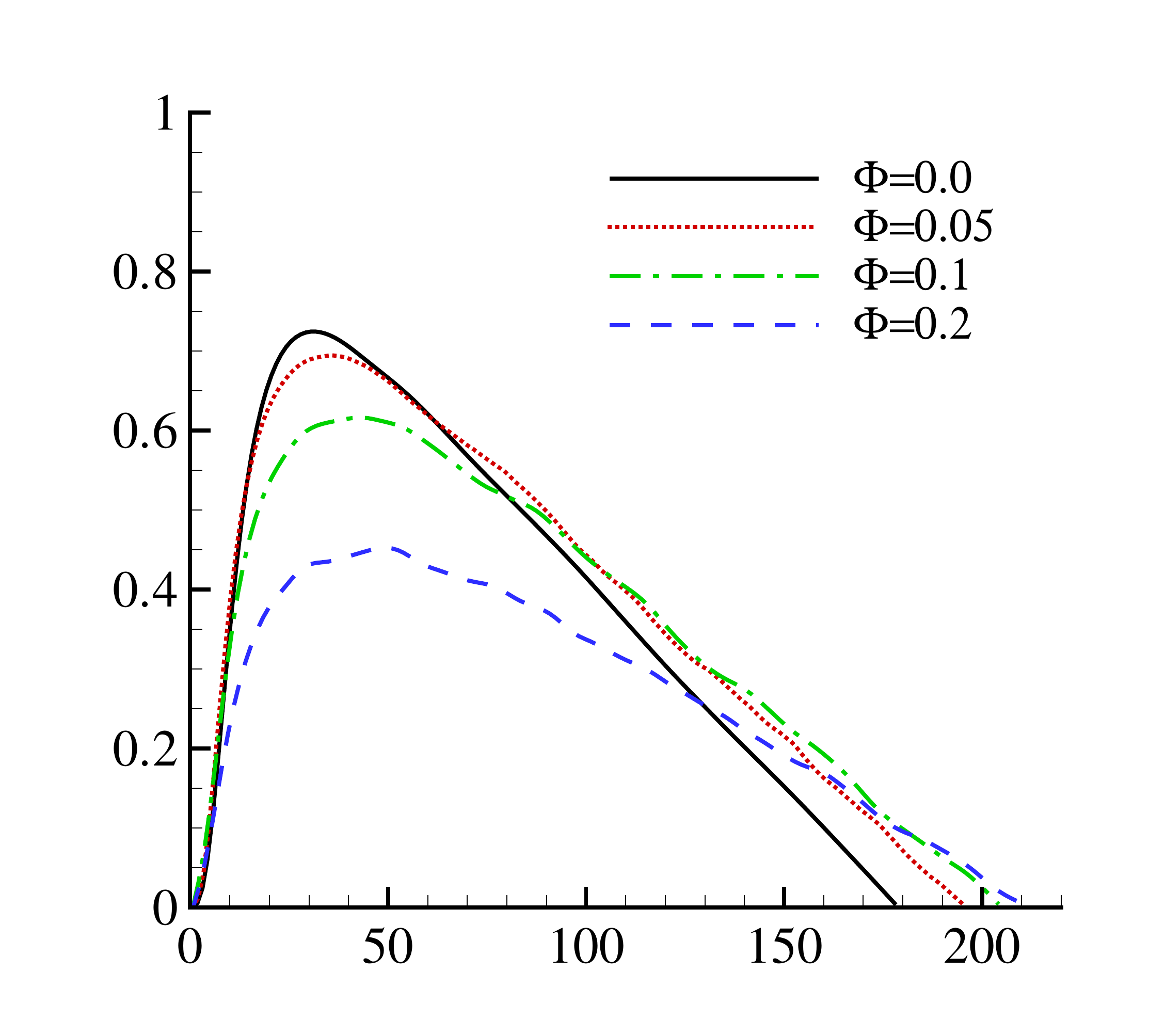} 
\put(-30,40){$d)$}\put(-90,3){$y^+$}\put(-185,78){\rotatebox{90}{$\langle u_f' v_f'\rangle^+$}}
\caption{\label{fig:4} Intensity 
of the fluctuation velocity components
and Reynolds shear stress for the fluid phase in inner units for different volume fraction
$\Phi$. Panel $a)$ streamwise ${u_f'}^+_{rms}$; $b)$ wall-normal ${v_f'}^+_{rms}$;
 and $c)$ spanwise ${w_f'}^+_{rms}$ velocity component;
  $d)$ shear-stress $\langle u_f' v_f'\rangle^+ $.
}
 \end{center}
 \end{figure} 

Further insight into the near wall dynamics can be gained by displaying the same quantities scaled in
inner units, see figure~\ref{fig:4}. 
The peak of the fluctuation intensity reduces for all the velocity components when divided by the friction velocity with the
only exception of the spanwise component.  More importantly, we observe that the fluctuation level monotonically
increases with $\Phi$ in the viscous sublayer.
This enhancement of the near-wall fluctuation can be explained by considering the 
squeezing motions occurring between
the wall and an incoming or outgoing particle. 
We also note that the peak of the Reynolds stresses decreases monotonically (when scaled by the friction velocity squared) becoming
about half of the expected value for the highest volume fraction considered here. 
The reduction of the Reynolds stress in inner units indicates that the increase of the drag is not due to an enhancement
of the turbulence activity, rather that it is linked to the solid phase dynamics.  
  
\begin{figure}
 \begin{center}
\includegraphics[width=.49\linewidth]{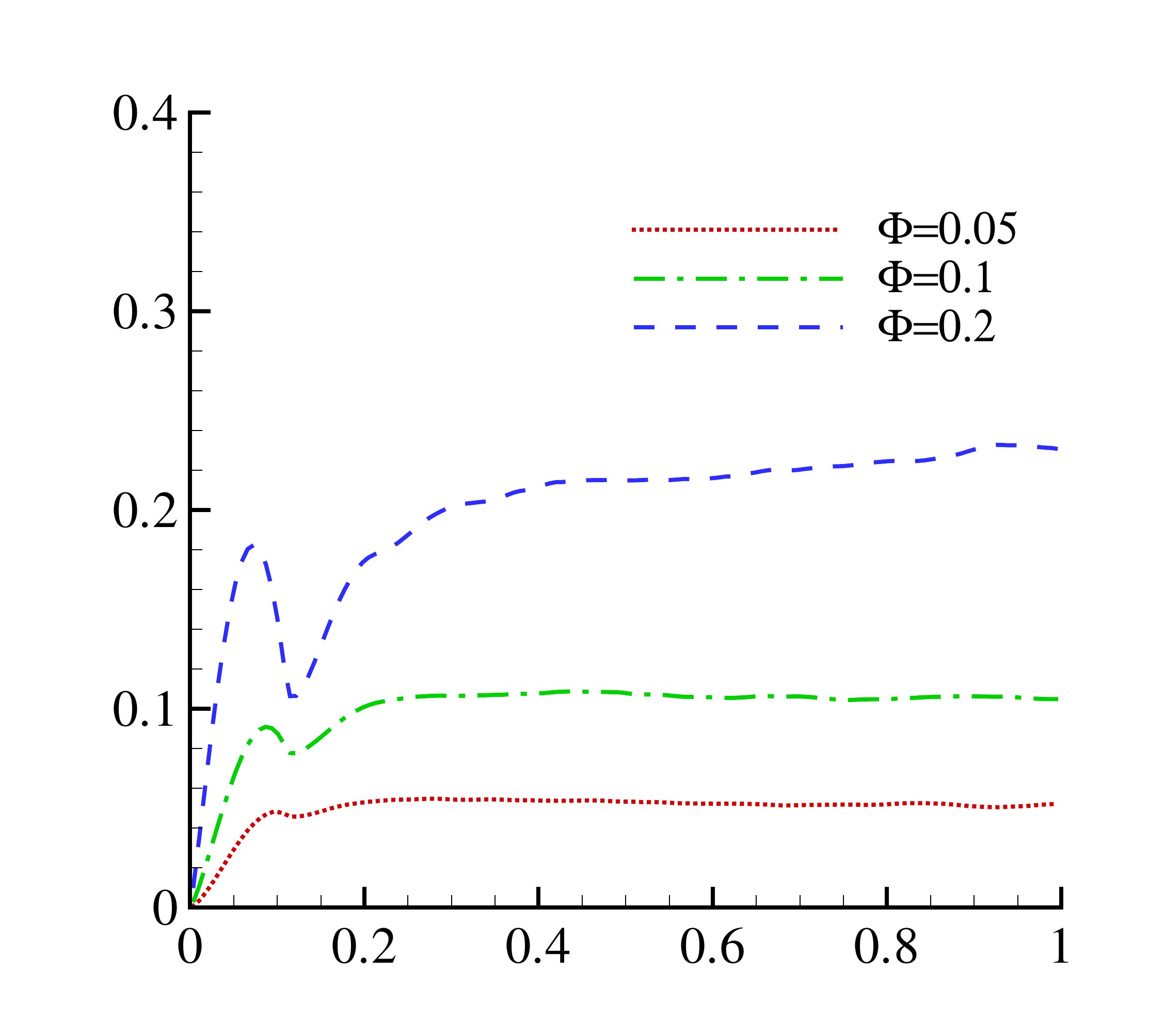} 
\put(-30,70){$a)$}\put(-90,3){$y$}\put(-185,82){\rotatebox{90}{$\phi$}}\hfill
 \includegraphics[width=.49\linewidth]{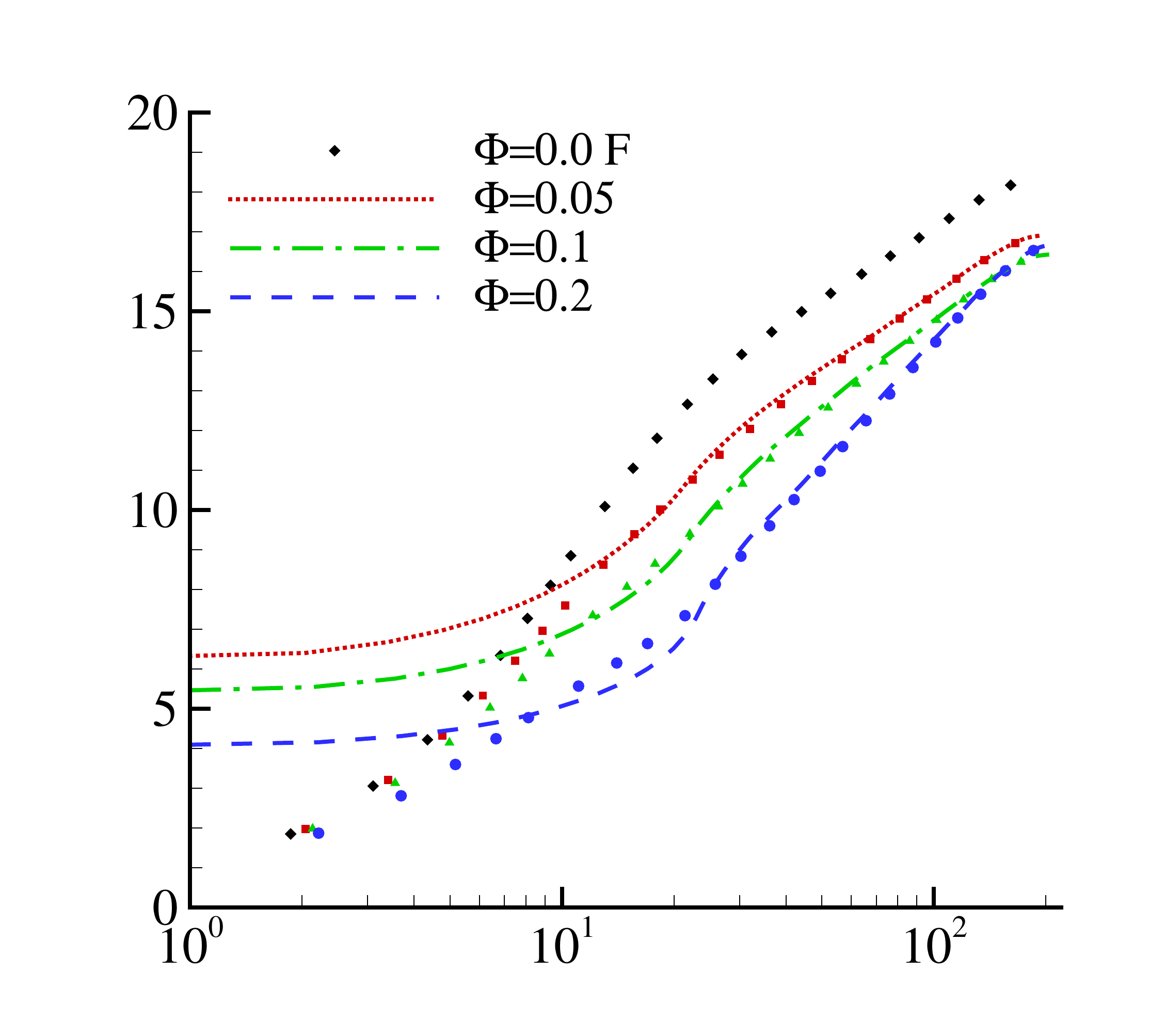} 
\put(-30,70){$b)$}\put(-95,3){$y^+$}\put(-185,82){\rotatebox{90}{$U_p^+$}}
\caption{\label{fig:5} Particle average data for different nominal   volume
fractions $\Phi$. $a)$ mean local volume fraction $\phi$ versus the wall-normal coordinate $y$ {(maximum statistical error $\pm0.01$)};  $b)$ mean particle velocity profile, $U_p^+=U_p/U_*$   in viscous units
$y^+=y/\delta_*$ (lines) {(maximum statistical error $\pm0.25\,U_*$)}. The 
mean fluid velocity $U_f^+$ is also reported for comparison (symbols).
}
 \end{center}
 \end{figure} 

To analyse the solid phase behaviour,  we report the mean local volume fraction $\phi(y)$ and the mean particle velocity 
$U_p$ in figure~\ref{fig:5}. The mean local volume fraction, panel $a)$, shows a
first local maximum around $y=0.06-0.1$, a value slightly larger than one particle
radius ($y=1/18$). Increasing the bulk volume fraction $\Phi$ the intensity of the peak grows, while
 a local minimum appears at $y\sim d_p=h/9$. As also observed in dense laminar regimes~\citep{yeomax_jfm10}, 
 a particle layer forms at the wall and becomes more intense when increasing
the bulk volume fraction $\Phi$. It should be noted however that these near-wall maxima are
smaller or similar to the bulk concentration, hence they are not related to the turbophoretic drift typically observed in dilute suspensions when particles are heavier
than the fluid~\citep{reeks}. Instead, these near-wall layers are induced by  the planar symmetry of the wall and the excluded finite volume of the solid spheres. {We believe that the formation 
of this  particle layer follows a mechanics similar to that usually observed in laminar Poiseuille and Couette flows~\citep{yeomax_jfm10,yeo2011numerical,pic_etal_prl13}.
Once a particle reaches the wall the strong wall-particle lubrication interaction stabilizes the particle wall-normal position that 
is therefore mainly affected by the collisions with other particles. Hence, it becomes difficult for the particles belonging
to the first layer to escape from it.     }
Figure~\ref{fig:5}$b)$  depicts the mean particle velocity $U^+_p$ in inner units (solid lines) where the fluid velocity is also reported
with symbols for a close comparison. As shown in the figure, solid and fluid phases flow with the same mean velocity in the whole 
channel with the exception of the first particle layer near the wall, $y^+\le20$, where particles have a mean velocity larger than the surrounding fluid. It should be considered here that while the velocity at the wall is zero for the fluid, this is not the case for the solid phase as particles can have a relative tangential motion.      
 
 \begin{figure}
 \begin{center}
 \includegraphics[width=.49\linewidth]{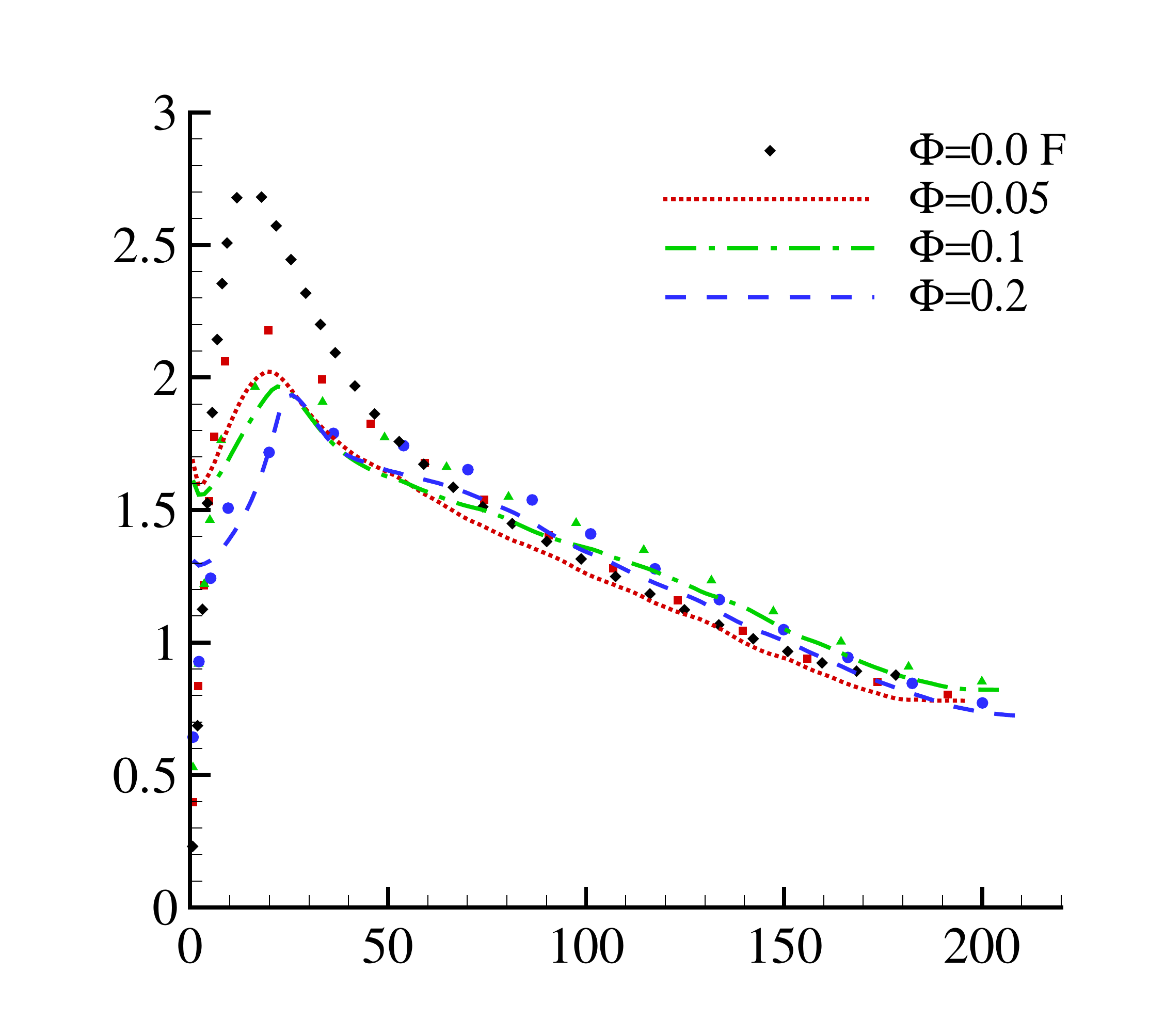} 
\put(-30,40){$a)$}\put(-90,3){$y^+$}\put(-185,80){\rotatebox{90}{${u_p'}^+_{rms}$}}\hfill
 \includegraphics[width=.49\linewidth]{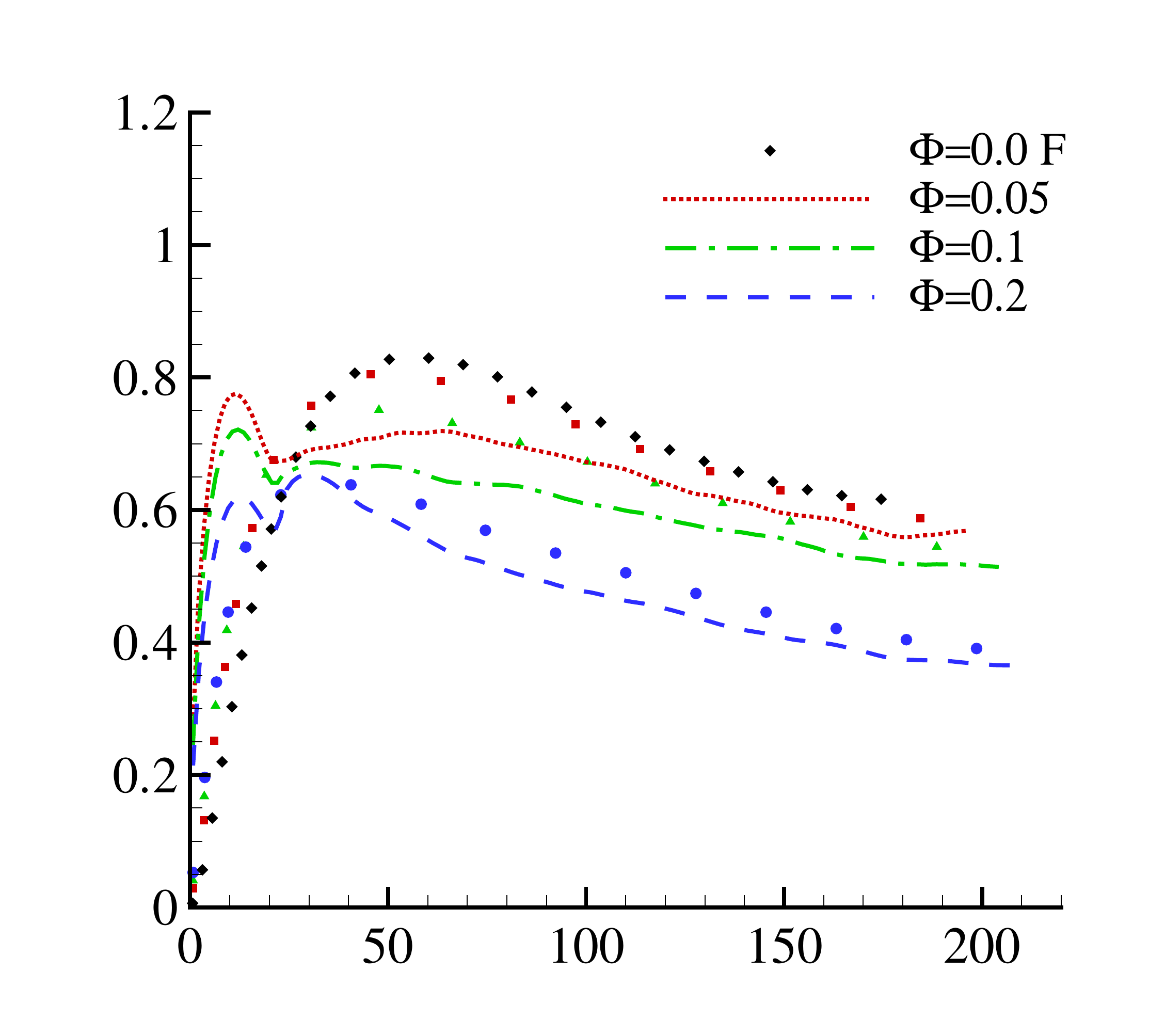} 
\put(-30,40){$b)$}\put(-90,3){$y^+$}\put(-185,80){\rotatebox{90}{${v_p'}^+_{rms}$}}\\[.1cm]
 \includegraphics[width=.49\linewidth]{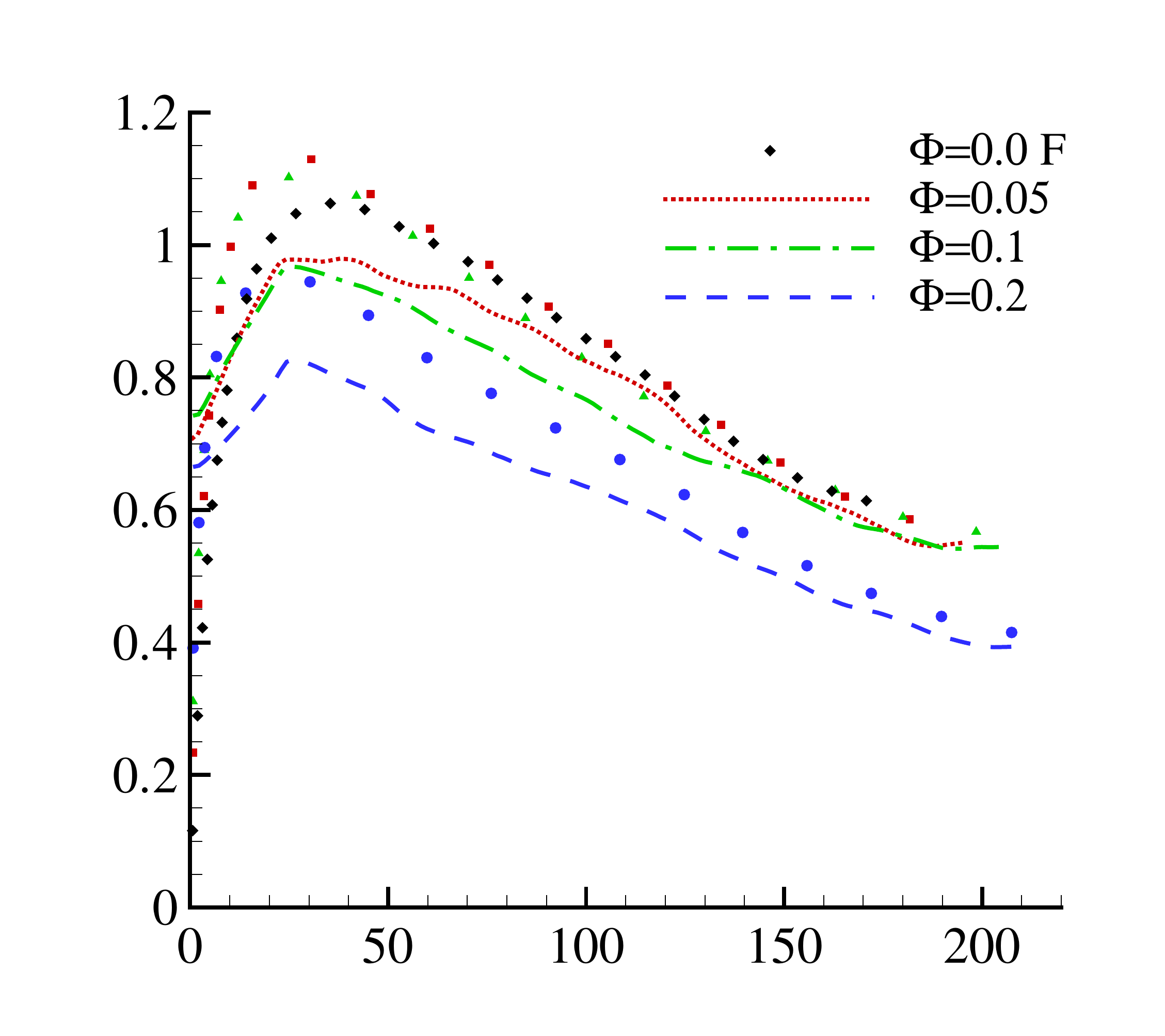} 
\put(-30,40){$c)$}\put(-90,3){$y^+$}\put(-185,80){\rotatebox{90}{${w_p'}^+_{rms}$}}\hfill
 \includegraphics[width=.49\linewidth]{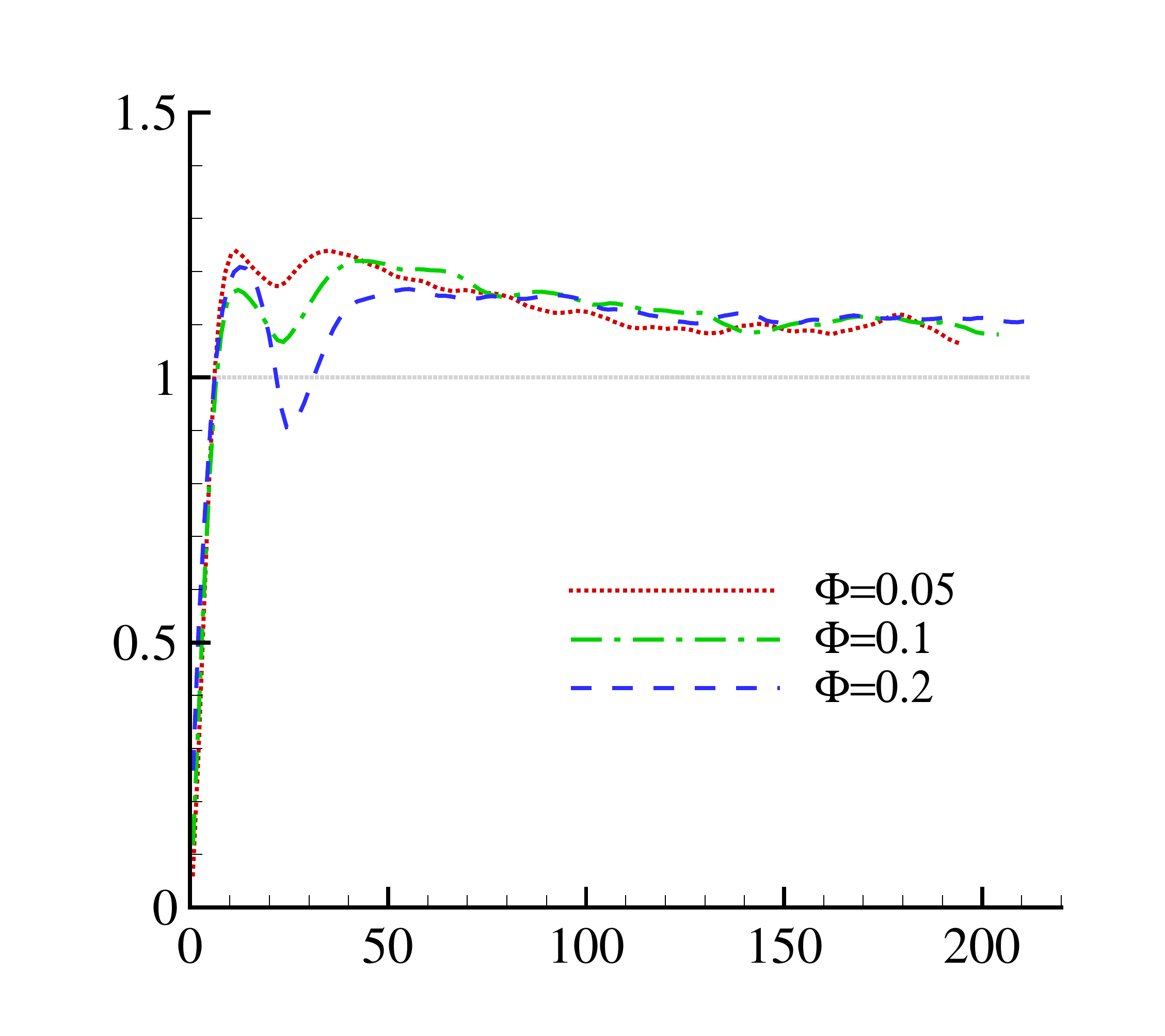} 
\put(-30,40){$d)$}\put(-90,3){$y^+$}\put(-180,78){\rotatebox{90}{$K_f/K_p$}}
\caption{\label{fig:6} Intensity of the fluctuation velocity components
for the solid phase in inner units for the different volume fraction
$\Phi$ studied {(maximum statistical error $\pm0.06\,U_*$)}. $a)$ streamwise ${u_p'}^+_{rms}$; $b)$ wall-normal ${v_p'}^+_{rms}$; and
 $c)$ spanwise ${w_p'}^+_{rms}$ component. Symbols represent the fluctuation levels of the fluid phase.
 Panel $d)$ displays the wall-normal profile of the ratio between the turbulent kinetic energy of the fluid and of the solid phase. 
}
 \end{center}
 \end{figure} 

The fluctuation intensities, rms, of the particle velocities are shown in figure~\ref{fig:6}, panels $a), b), c)$, in inner units. 
The streamwise component ${u'_p}_{rms}$ shows similar fluctuation levels for both
phases and all $\Phi$ with some small differences close to the wall where the solid phase fluctuations do not
vanish. Considering the three velocity components we generally observe that particles tend to fluctuate less than the
fluid at the same position except for the region close to the wall. This behaviour is summarised in the panel $d)$ of the same figure
where we display the ratio between the turbulent kinetic energy of the fluid and of the solid phase, 
$K_f/K_p=(u_f'^2+v_f'^2+w_f'^2)/(u_p'^2+v_p'^2+w_p'^2)$. 
Besides a thin region
close to the wall, the fluid turbulent kinetic energy is higher than the energy of the solid phase by about $10\div20\%$. 
The higher particle fluctuation level in the near wall region, due to the absence of a no-slip condition at the wall, suggests that
this is the cause of the near-wall enhancement of the fluid fluctuation level (compared to the single phase flow) discussed above.
One last remark concerns the local peak of the wall-normal particle velocity fluctuation close to wall.
This maximum originates from particles that reach and leave the first layer at the wall. 
In this region the fluid velocity fluctuations increase with $\Phi$, though the maximum
for the solid phase decreases. This is not contradictory, as it just indicates that at small volume fractions the incoming/leaving 
particles are fewer, but faster; increasing $\Phi$ more particles enter and leave the first layer though at smaller velocity as it is more crowdy. 

{
Figure~\ref{fig:rot} reports the mean particle angular velocity $\Omega_z$, panel $a)$, and the particle angular velocity fluctuation
 rms in the spanwise 
$\omega'_{z\, rms}$, panel $b)$, streamwise 
$\omega'_{x\, rms}$, panel $c)$ and wallnormal $\omega'_{y\, rms}$, panel $d)$, directions. 
The mean particle angular velocity $\Omega_z$ is maximum close to the wall and vanishes in the
centerline for symmetry. This behavior indicates that the particle belonging to the layer close to the wall tend to 
roll on the wall minimizing their local slip velocity, which as previously discussed is in principle not vanishing. 
The slight reduction of the maximum rotation observed when increasing the volume fraction $\Phi$ is induced by the more intense particle-particle  interactions occurring in the first layer. Interestingly, at $\Phi=0.2$, in the bulk of the flow, 
the mean angular velocity is higher than in the other cases. This can be explained by  the higher fluid velocity
gradient exhibited in this region at $\Phi=0.2$, see for instance Figure~\ref{fig:2}. 
Concerning the fluctuation levels of the particle angular velocity, we note that the maximum of each  component occurs  
near the wall showing values that are about  $15\div25\%$ of the mean value.  Near the peak, the spanwise 
fluctuating component shows higher intensity $\omega'_{z\, rms}$ driven by the inhomogeneity of the mean angular velocity,
while the three components become of similar magnitude near the centerline (isotropy). The densest case shows slightly smaller
fluctuations whereas the flows at $\Phi=0.05;0.1$ exhibit almost the same values.}
  
  \begin{figure}
 \begin{center}
 \includegraphics[width=.49\linewidth]{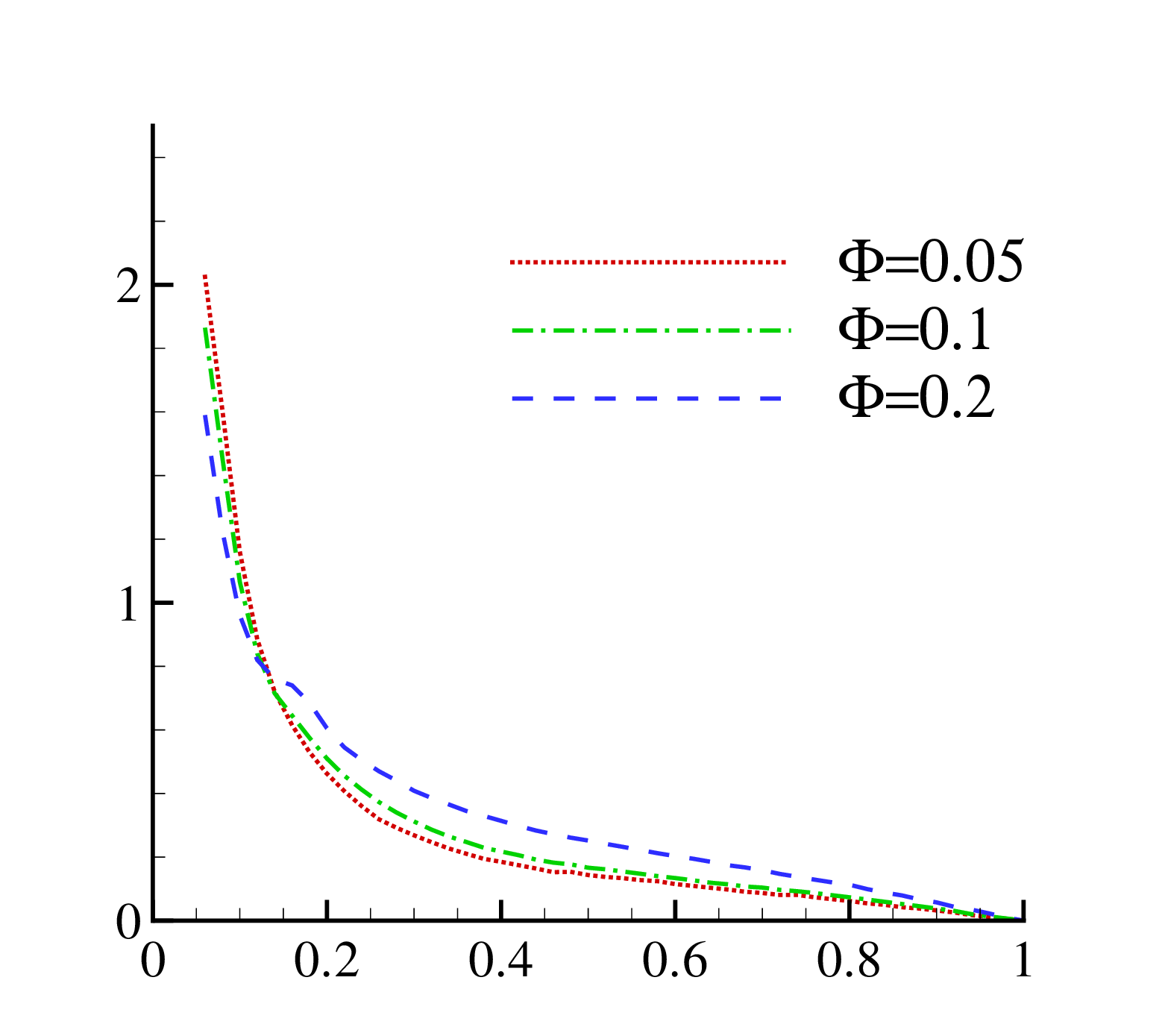} 
\put(-30,70){$a)$}\put(-97,3){$y$}\put(-185,80){\rotatebox{90}{$\Omega_z$}}\hfill
 \includegraphics[width=.49\linewidth]{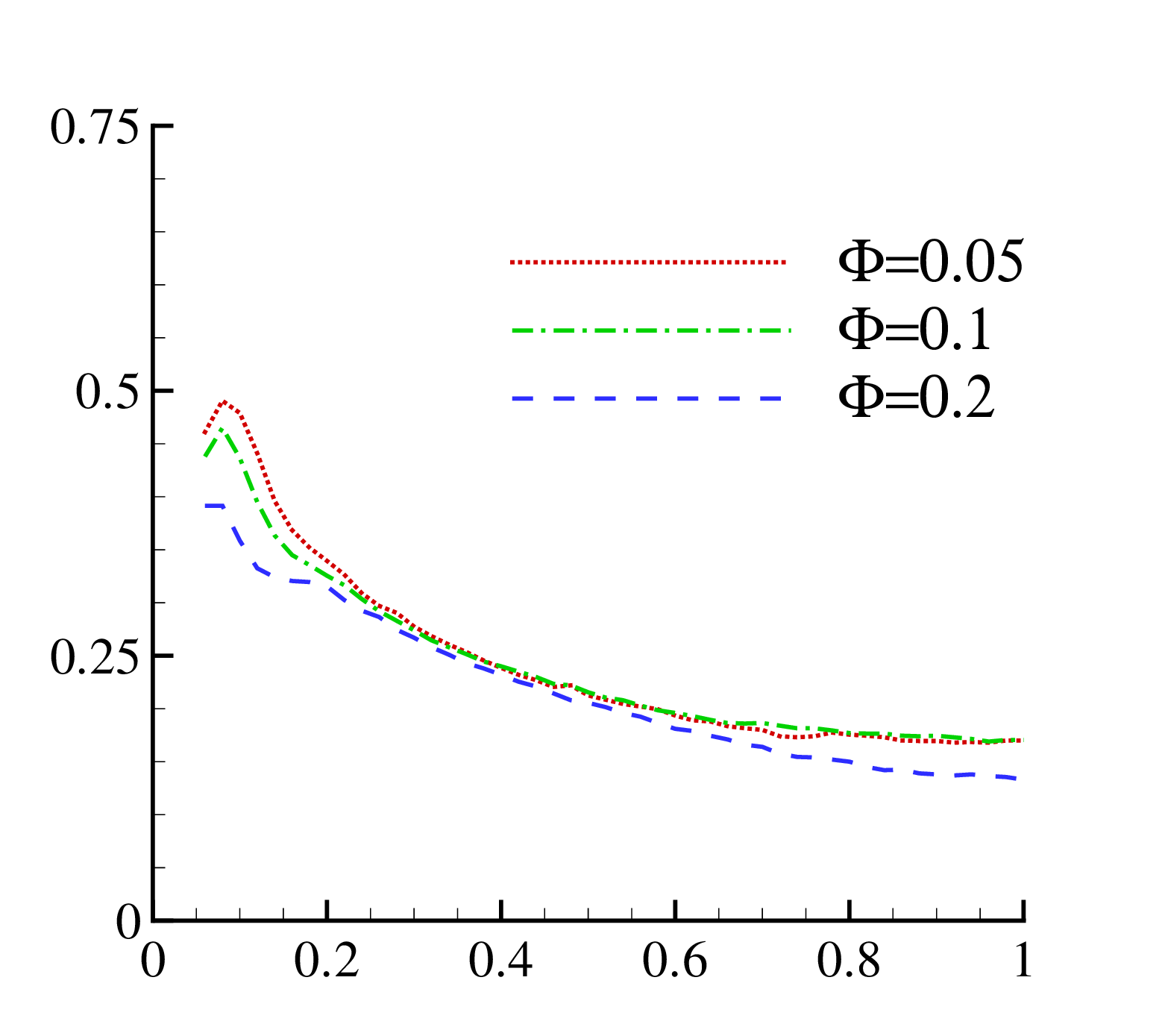} 
\put(-30,70){$b)$}\put(-97,3){$y$}\put(-190,80){\rotatebox{90}{$\omega'_{z\, rms}$}}\\[.1cm]
 \includegraphics[width=.49\linewidth]{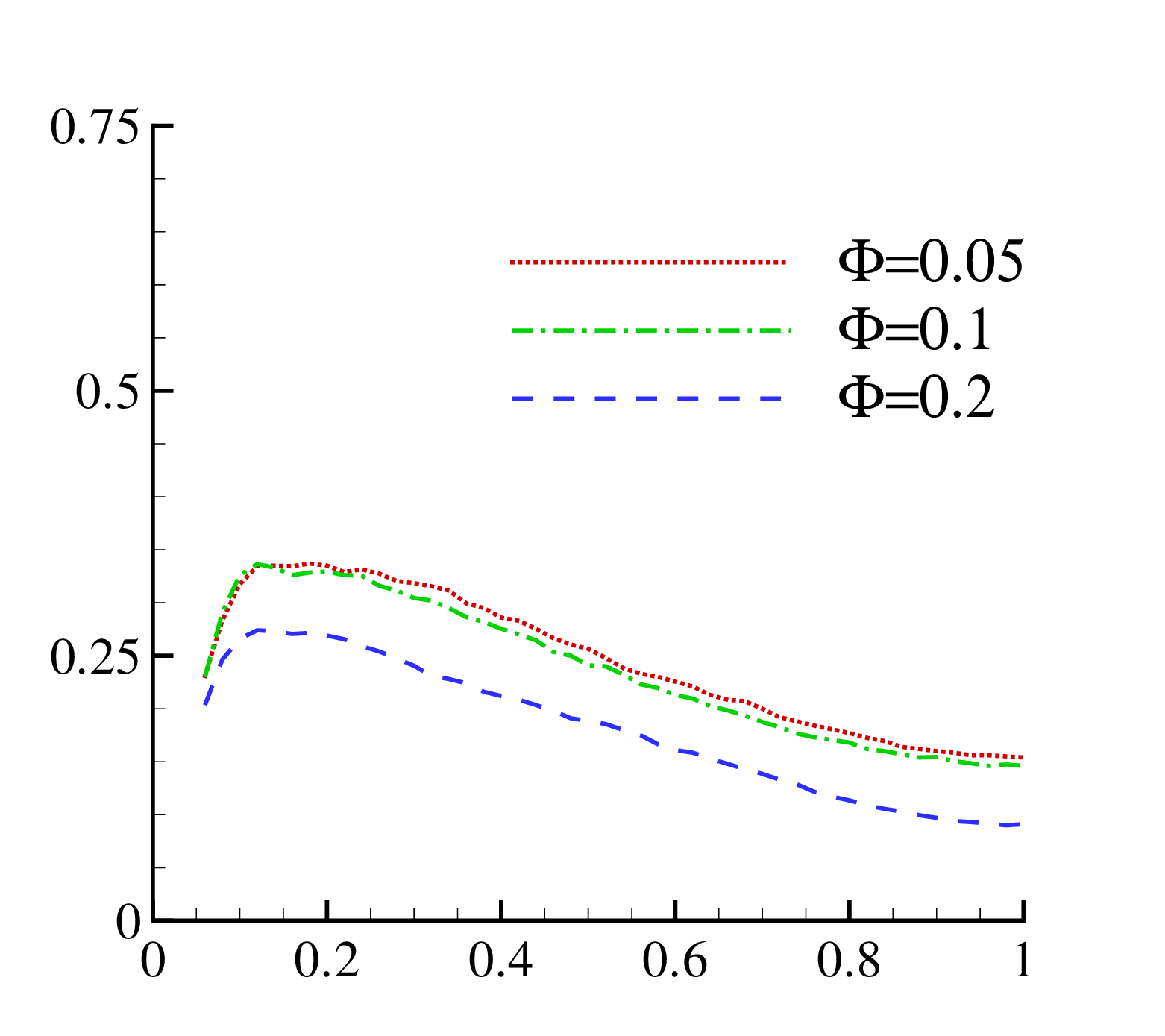} 
\put(-30,70){$c)$}\put(-97,3){$y$}\put(-190,80){\rotatebox{90}{$\omega'_{x\, rms}$}}\hfill
 \includegraphics[width=.49\linewidth]{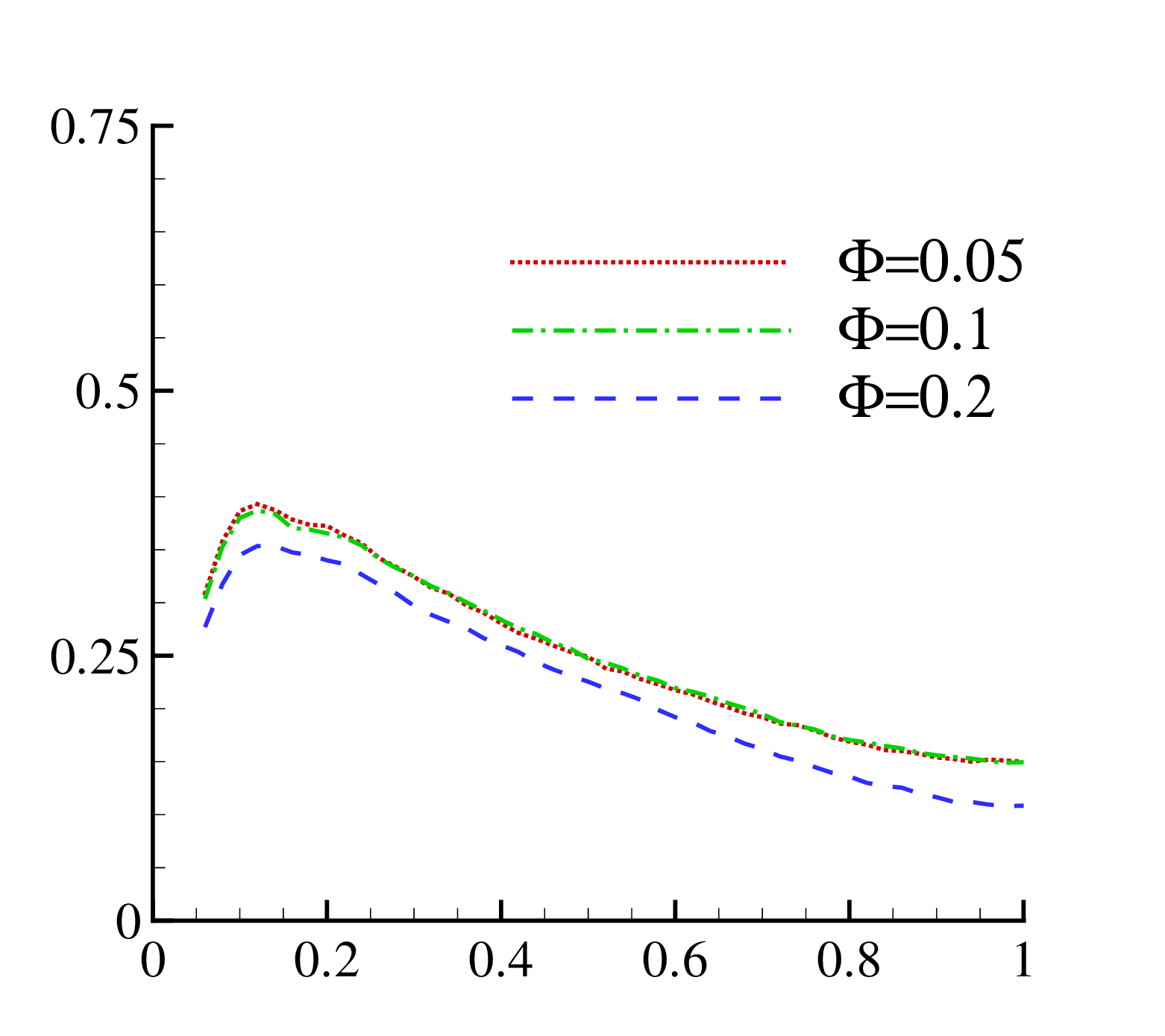} 
\put(-30,70){$d)$}\put(-97,3){$y$}\put(-190,78){\rotatebox{90}{$\omega'_{y\, rms}$}}
\caption{\label{fig:rot} {Particle angular velocity statistics in outer units for the different volume fractions
$\Phi$ studied. $a)$ mean angular velocity (spanwise component) $\Omega_z$; 
$b)$ fluctuation rms of the spanwise fluctuation component $\omega'_{z\, rms}$;  
$c)$ fluctuation rms of the streamwise component $\omega'_{x\, rms}$;
$d)$ fluctuation rms of the wall-normal component $\omega'_{y\, rms}$. }
}
 \end{center}
 \end{figure}

 \begin{figure}
 \begin{center}
 \includegraphics[width=.49\linewidth]{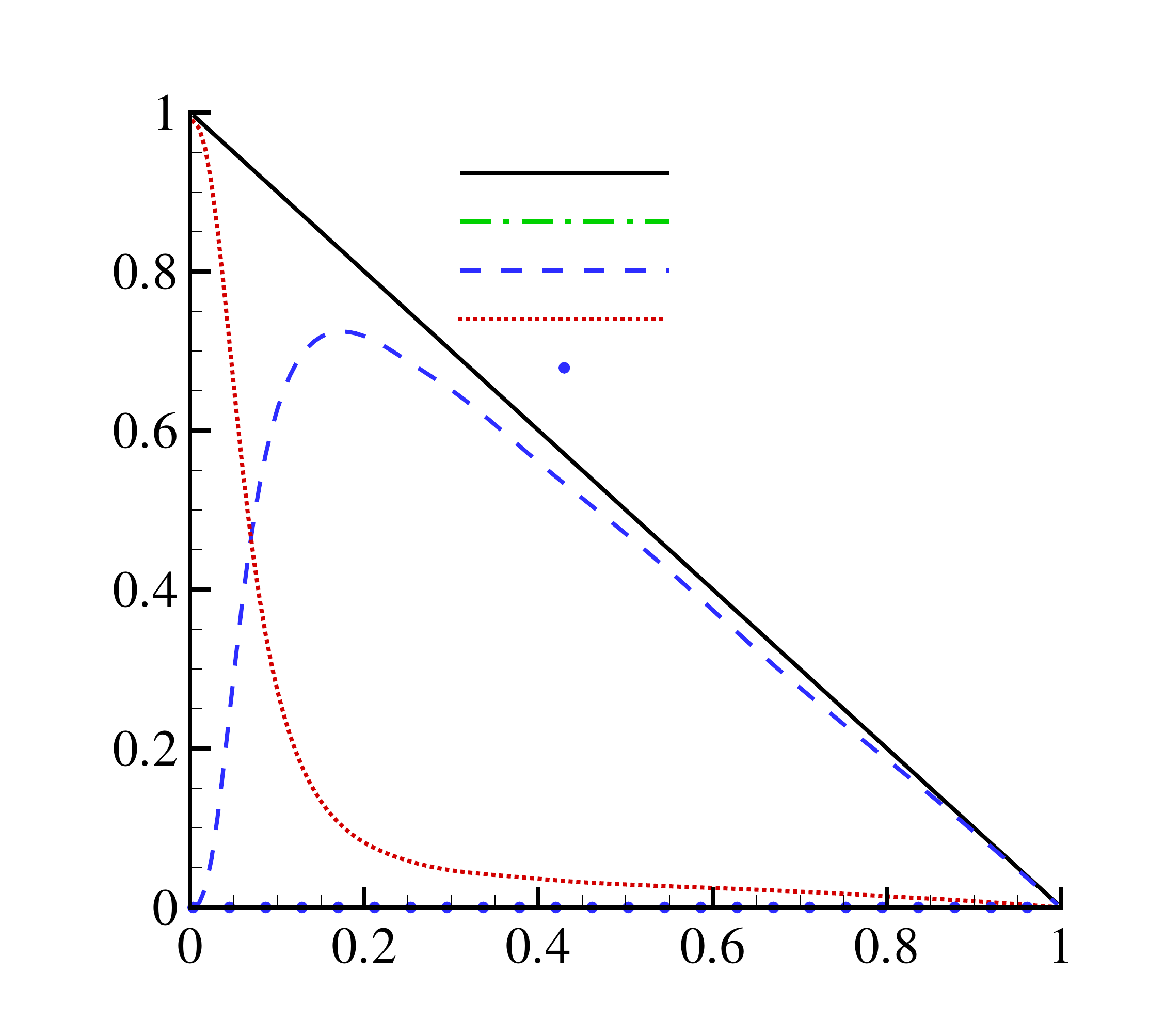} 
\put(-50,70){$\Phi=0$}\put(-90,3){$y$}\put(-185,80){\rotatebox{90}{$\tau_i^+$}}
{\scriptsize \put(-77,137.5){$\tau$}\put(-77,129.5){$\tau_P$}\put(-77,121.5){$\tau_T$}
\put(-77,113.5){$\tau_{V}$}\put(-77,105.5){$\tau_{T_p}$}}
\hfill
 \includegraphics[width=.49\linewidth]{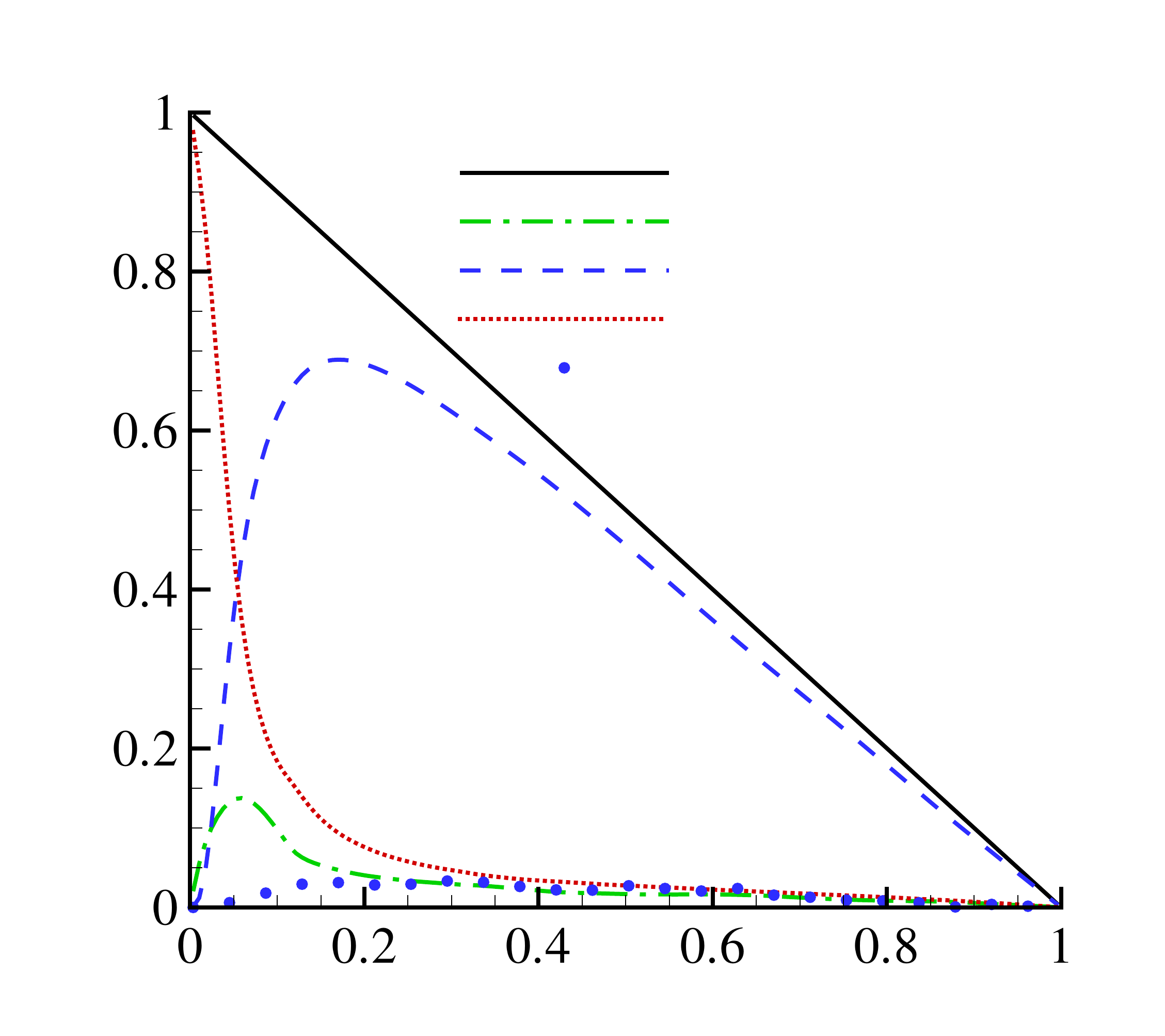} 
\put(-50,70){$\Phi=0.05$}\put(-90,3){$y$}\put(-185,80){\rotatebox{90}{$\tau_i^+$}}
{\scriptsize \put(-77,137.5){$\tau$}\put(-77,129.5){$\tau_P$}\put(-77,121.5){$\tau_T$}
\put(-77,113.5){$\tau_{V}$}\put(-77,105.5){$\tau_{T_p}$}}
\\[.1cm]
 \includegraphics[width=.49\linewidth]{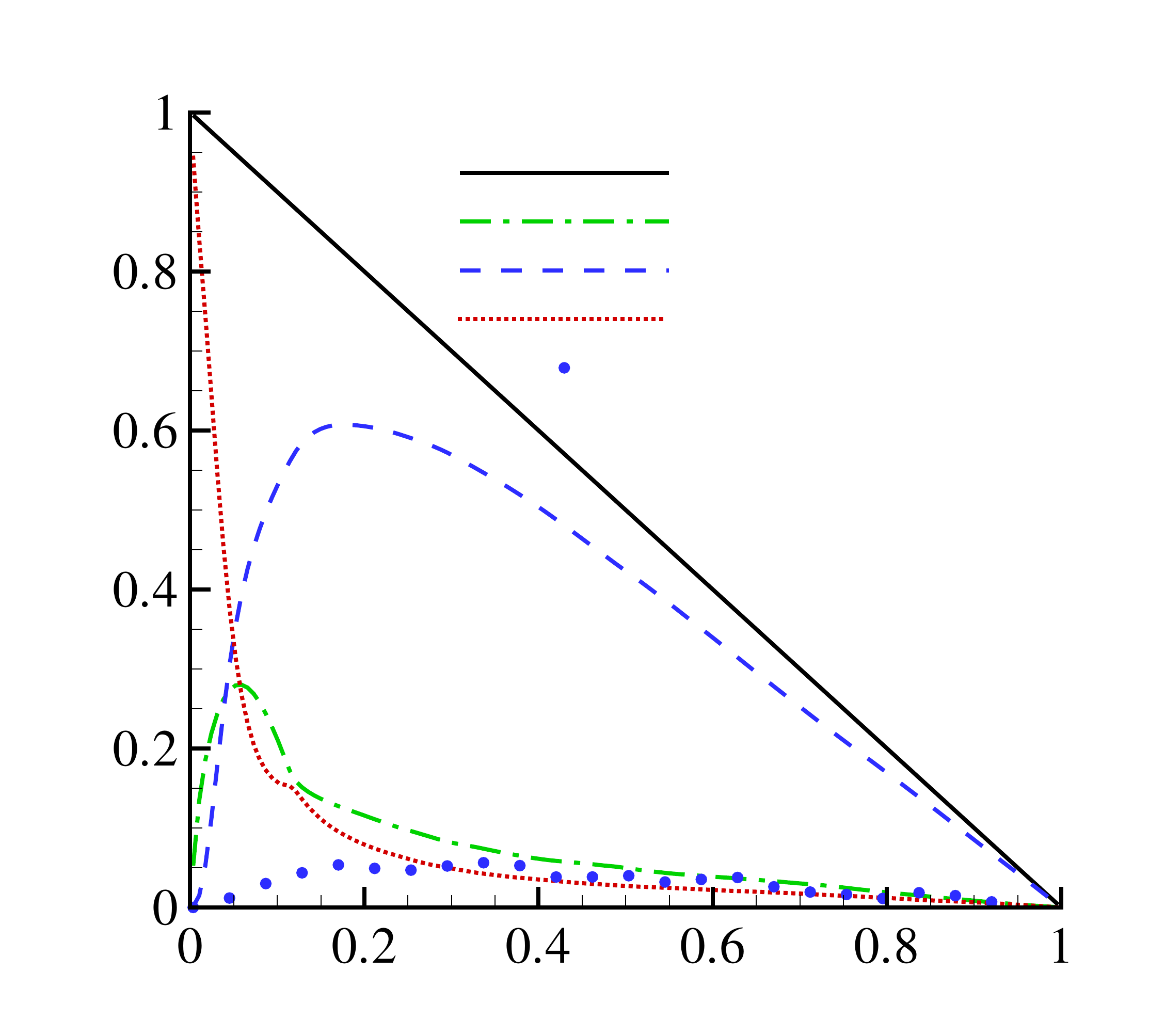} 
\put(-50,70){$\Phi=0.1$}\put(-90,3){$y$}\put(-185,80){\rotatebox{90}{$\tau_i^+$}}
{\scriptsize \put(-77,137.5){$\tau$}\put(-77,129.5){$\tau_P$}\put(-77,121.5){$\tau_T$}
\put(-77,113.5){$\tau_{V}$}\put(-77,105.5){$\tau_{T_p}$}}\hfill
 \includegraphics[width=.49\linewidth]{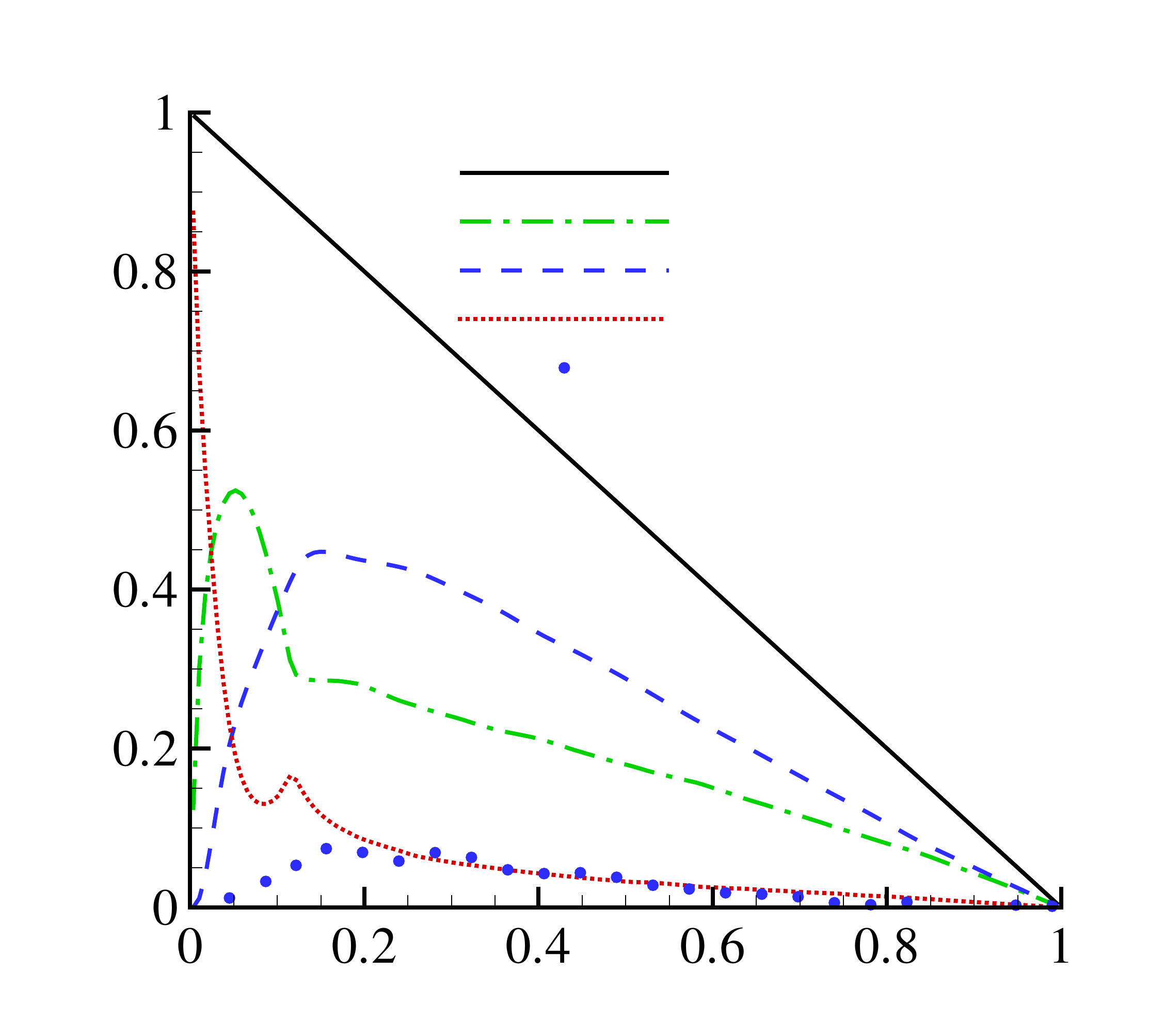} 
\put(-50,70){$\Phi=0.2$}\put(-90,3){$y$}\put(-185,80){\rotatebox{90}{$\tau_i^+$}}
{\scriptsize \put(-77,137.5){$\tau$}\put(-77,129.5){$\tau_P$}\put(-77,121.5){$\tau_T$}
\put(-77,113.5){$\tau_{V}$}\put(-77,105.5){$\tau_{T_p}$}}
\caption{\label{fig:7} Momentum budget for the different bulk volume fractions $\Phi$ under investigation. The wall is at $y=0$, whereas $y=1$ is the channel centreline. $\tau_V$, $\tau_T$ and $\tau_P$ represent the viscous, turbulent and particle induced stresses. $\tau_{T_p}$ is the particle Reynolds stress and $\tau=U_*^2\,(1-y)$ the total stress.}
 \end{center}
 \end{figure} 

\subsection{Total stress balance}
\label{sec:bud}
The understanding of the  momentum exchange between the two phases in dense particle-laden
turbulent channel flows is conveniently addressed by examining the streamwise momentum budget, i.e.\ the
average stress budget. Following the rationale on the mean momentum balance given in
appendix \S~\ref{app}, see also \cite*{mar_etal_ijmf99} and \cite{zhapro_etal_pf10} for more details, we can write the whole budget as the sum of three terms:
\begin{equation}
\tau=\tau_V+\tau_T+\tau_P\, ,
\label{eq:stress}
\end{equation}
where $\tau=U_*^2\,(1-y)$ is the total stress, $\tau_V=\nu(1-\phi)(d U_f)/(dy)$
is the viscous stress, $\tau_T=-\langle u'_c v'_c \rangle$ is the turbulent Reynolds shear
stress of the combined phase $\langle u'_c v'_c \rangle=
\phi \langle u'_p v'_p \rangle+(1-\phi)\langle u'_f v'_f \rangle$ 
(with the particle Reynolds stress $\phi \langle u'_p v'_p \rangle=\tau_{T_p}$)
and $\tau_P= (\phi/\rho)(\langle \sigma_{p\,xy} \rangle)$ the particle induced stress.


Figure~\ref{fig:7} reports the stress balance given in eq.~\eqref{eq:stress} from the simulations for the four bulk volume fractions $\Phi$
presented here and normalised by the corresponding friction
velocity squared, $U_*^2$ \red{(the particle induced stress has been indirectly calculated from the balance)}. 
As already known for the single phase flow~\citep{pope2000turbulent}, the total stress $\tau$ 
is mainly given by the turbulent Reynolds stress term 
for $y \ge 0.2$. 
The relevance of the viscous stress increases \lb{ approaching the wall}, becoming the leading
term as the Reynolds stress is zero at the wall. At $\Phi$=0.05, see figure~\ref{fig:7}$b)$, the basic picture remains unaltered with the particle induced stress $\tau_P$ 
showing a not negligible contribution only near the wall; note that the particle turbulent Reynolds
stress is still negligible in this configuration. Increasing the volume fraction to $\Phi=0.1$, panel $c)$ in the figure, the particle-induced stress becomes of the same order of magnitude as
the other terms in the near wall region, $y \approx 0.05$, which roughly 
corresponds to a particle radius. The contribution from the particle stress, 
though still sub-leading with respect to the turbulent stress $\tau_T$,
is important throughout the whole channel. Note also that  that the turbulent stress associated to the solid
phase alone, $\tau_{T_p}$, amounts to $\sim10\%$ of the total $\tau_T$, scaling almost linearly with the volume fraction. 
For the highest volume fraction considered, $\Phi=0.2$, see panel $d)$, the near wall dynamics is dominated by the particle-induced stresses. This is now the leading term around $y=0.05$. Moreover the total stress in the bulk of the flow, $y>0.2$,
is \emph{transmitted} by the turbulent shear stress $\tau_T$ 
and the particle induced stress $\tau_P$ in similar shares. In other words, 
the turbulent shear stress amounts to about half of the total stress in the bulk of the flow. This indicates that 
the turbulent dynamics is strongly altered by the dense particle concentration: though the system is still turbulent,  
the particle-induced stress becomes crucial in transferring the mean stress through the channel. 

This behaviour is consistent with the decrease of the turbulence activity previously discussed for the flow with the highest particle number, $\Phi=0.2$. 
As mentioned  above, although the turbulence intensities and the Reynolds shear stress are attenuated, 
the total drag, i.e.\ the friction Reynolds number increases. One can therefore conclude that this increase of the total drag is not associated to a turbulence enhancement, but to an increase of the particle-induced stress, or borrowing rheological terms, to an increase of the effective viscosity of the flowing medium. 

In order to quantify the level of turbulence activity, we can define the \emph{turbulent} friction velocity as
\begin{equation}
U_*^T=\left.\sqrt{\frac{d\langle u'_c v'_c\rangle}{dy}}\right|_{y=1},
\end{equation}
that is the square root of the wall-normal derivative of the Reynolds stress profile at the centreline ($y=1$).
This quantity has been chosen because it can be shown that the turbulent friction velocity well approximates the wall friction velocity for unladen cases at high bulk Reynolds number,
$U_*^T=U_*+{\cal O}(1/Re)$, see e.g.~\cite{pope2000turbulent}. 

Figure~\ref{fig:8}$a)$ reports the turbulent
Reynolds stress of the combined phase $\langle u'_c v'_c \rangle$ in outer units together with a straight line indicating the slope at $y=1$. 
The intercept of this line originating at $(y,\langle u'_c v'_c\rangle)=(1,0)$ with the vertical axis provides the value of the 
 turbulent friction velocity, $U_*^T$ as defined above.
As clear from the figure, $U_*^T$ increases when adding the solid phase until $\Phi=0.1$ and then decreases at $\Phi=0.2$. 
Using these values, we can then define a \emph{turbulent} friction Reynolds number: $Re_T= U_*^T\,h/\nu$. 
Since $Re_T$ is proportional to $U_*^T$, it follows that $Re_T=Re_\tau+ {\cal O}(1/Re)$ for high Reynolds number single-phase turbulent 
channel flows.
The panel $b)$ of figure~\ref{fig:8} depicts the friction Reynolds number $Re_\tau$ and the turbulent Reynolds number $Re_T$ just introduced  versus the bulk volume fraction $\Phi$. 
The values of the two Reynolds numbers in the unladen case, $\Phi=0$, are close, as expected. Increasing the 
volume fraction, both $Re_\tau$ and $Re_T$ increase up to $\Phi=0.1$, $Re_T$ at a slower rate. 
Interestingly, at $\Phi=0.2$, the turbulent friction Reynolds number $Re_T$ suddenly decreases, 
whereas the friction Reynolds number based on the actual wall-shear still increases. 

The friction velocity and Reynolds number are a measure of the overall drag as they are proportional to the imposed pressure gradient, while the turbulent friction velocity and corresponding Reynolds number introduced here indicate only the portion of the drag directly induced by the turbulent activity. We therefore conclude
that in dense cases, i.e. $\Phi=0.2$, a turbulent drag reduction indeed occurs and this is related to a reduced turbulence activity. 
Nonetheless, this turbulent drag reduction does not reflect in a decrease of the total drag at the Reynolds number
 investigated here because the particle-induced stress (increased viscosity of the suspension) more than counteracts the positive effect due to the reduced turbulent mixing. 
The observations emerging from our analysis of the momentum budget explain and are consistent with the large reduction of the von K{\'a}rm{\'a}n constant $k$ found for this dense case and reported in table~\ref{tab:1}.

 \begin{figure}
 \begin{center}
 \includegraphics[width=.49\linewidth]{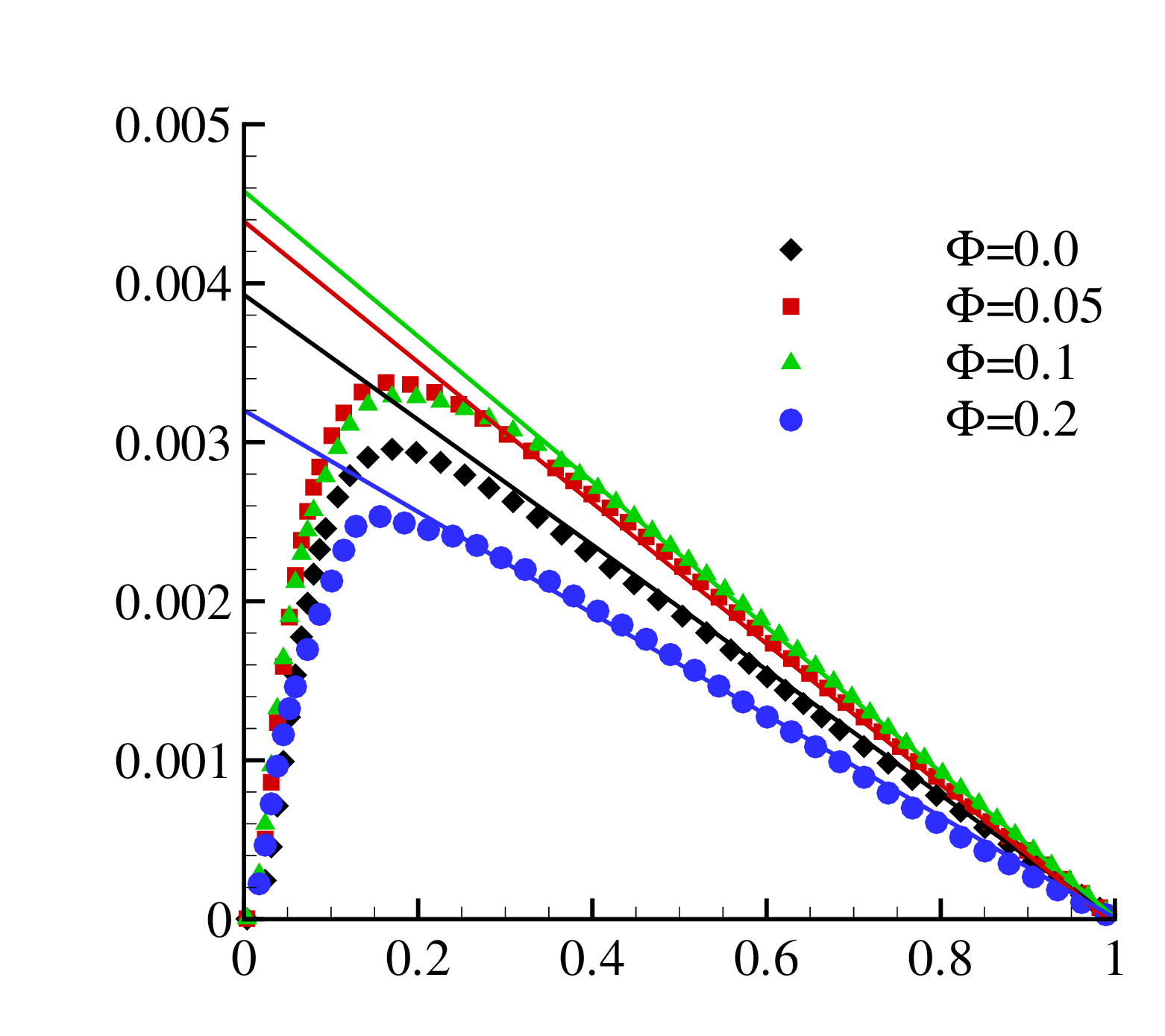} 
\put(-30,70){$a)$}
\put(-90,3){$y$}\put(-185,80){\rotatebox{90}{$-\langle u'_C v'_C\rangle$}}
\hfill
 \includegraphics[width=.49\linewidth]{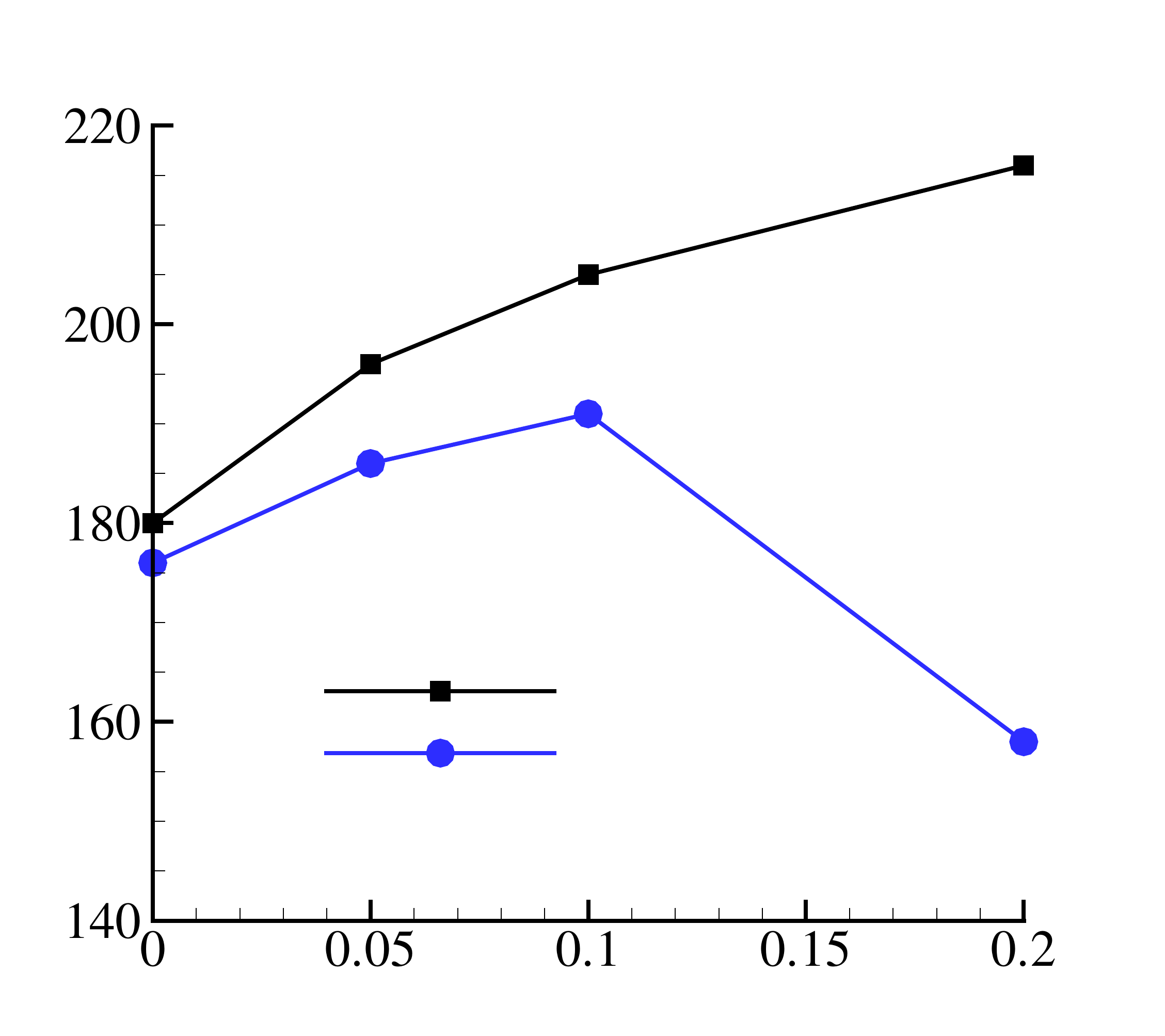} 
\put(-30,70){$b)$}
\put(-90,3){$\Phi$}\put(-188,65){\rotatebox{90}{$Re_\tau,\,Re_T$}}
{\small \put(-93.5,53.2){$Re_\tau$}}
{\small \put(-93.5,43.2){$Re_T$}}
\caption{\label{fig:8} $a)$: wall-normal profiles of the shear stress of the combined phase $\langle u'_C v'_C \rangle/U_0^2$ 
(symbols) and linear fitting of the slope of the profile at the centreline, $y=1$, depicted with solid lines.
$b)$: Friction Reynolds number $Re_\tau=u_* h/\nu$  and Turbulent Friction Reynolds number $Re_T=U_*^T h/\nu$
versus the bulk volume fraction $\Phi$ from the simulations presented here.}
 \end{center}
 \end{figure} 

\subsection{Velocity correlations}

Further understanding of the effect of the solid phase on the turbulent channel flow is obtained by examining the two-point spatial correlation
of the velocity field. It is well known that
the auto-correlations of the streamwise and wall-normal velocity along the spanwise direction,
\begin{align}
&R_{uu}(y,\Delta z)=\frac{\langle u'(x,y,z,t)u'(x,y,z+\Delta z,t)\rangle}{{u'}^2_{rms}}\, ,\\
&R_{vv}(y,\Delta z)=\frac{\langle v'(x,y,z,t)v'(x,y,z+\Delta z,t)\rangle}{{v'}^2_{rms}}
\end{align}
show a negative minimum value in the near wall region 
around $\Delta z^+=60\div80$ and $\Delta z^+=30\div40$, respectively, for a single-phase turbulent flow. 
 
\sloppy
These values reflect
the typical structures of wall-bounded turbulence, i.e.\
quasi-streamwise vortices and low speed streaks that sustain the turbulence process
\citep{pope2000turbulent,kmm,waleffe1997self,Brandt201480}. It has also been observed that in drag reducing turbulent flows 
the width and the spacing of these characteristic structures increases \citep{stone2002toward,de2002dns}, leading
to an increase of the spanwise separation of these minima.

 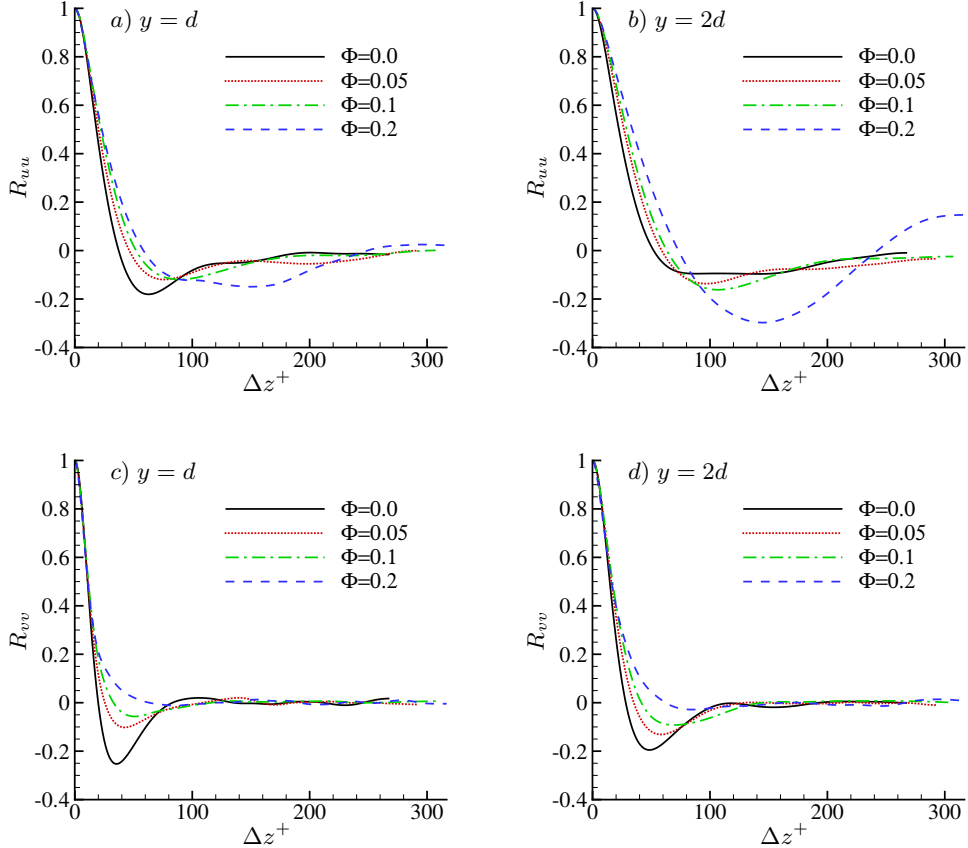
\begin{figure}
 \begin{center}
 \includegraphics[width=.49\linewidth]{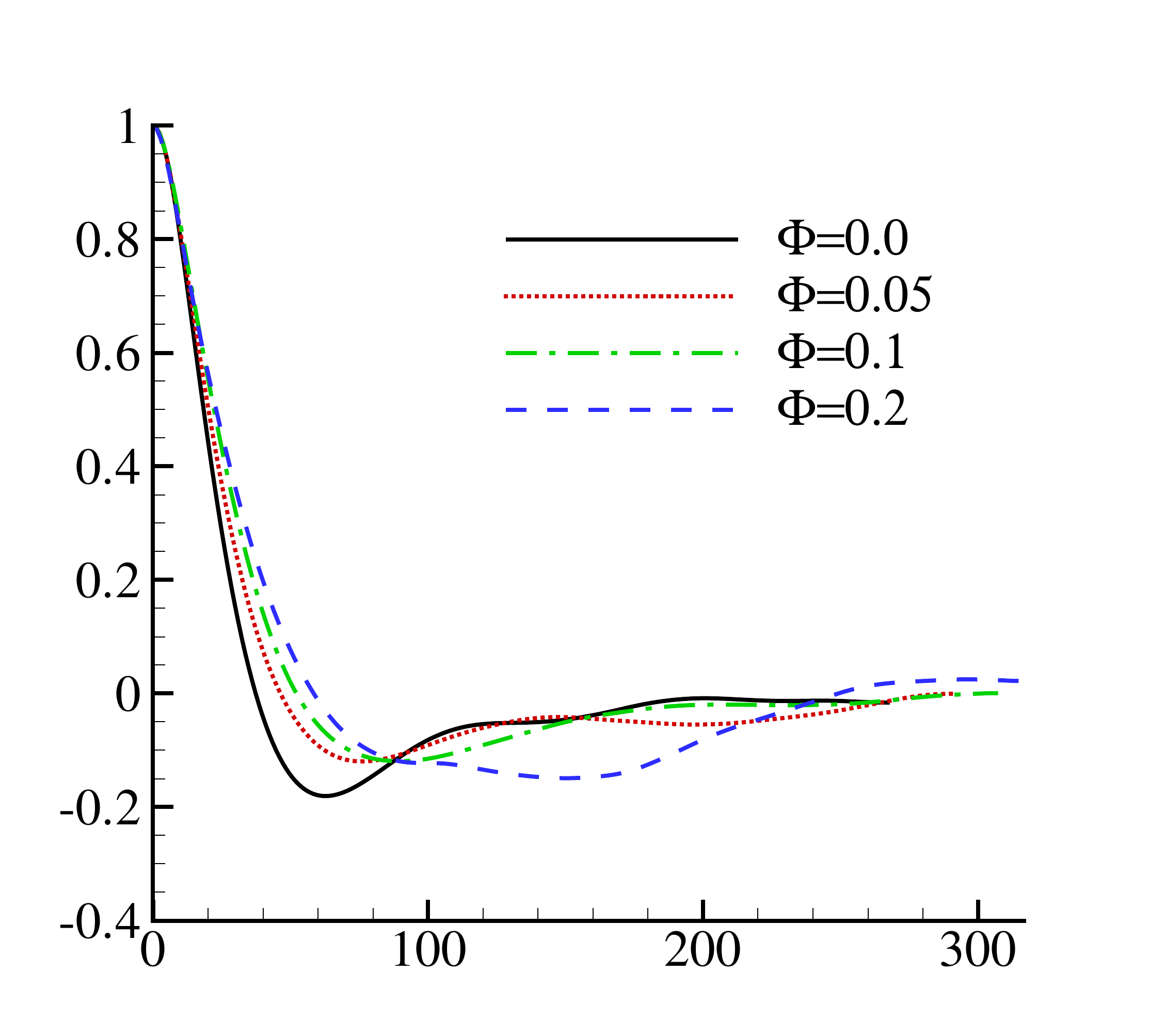} 
\put(-150,140){$a)\;y=d$}\put(-100,2){$\Delta z^+$}\put(-188,75){\rotatebox{90}{$R_{uu}$}}\hfill
 \includegraphics[width=.49\linewidth]{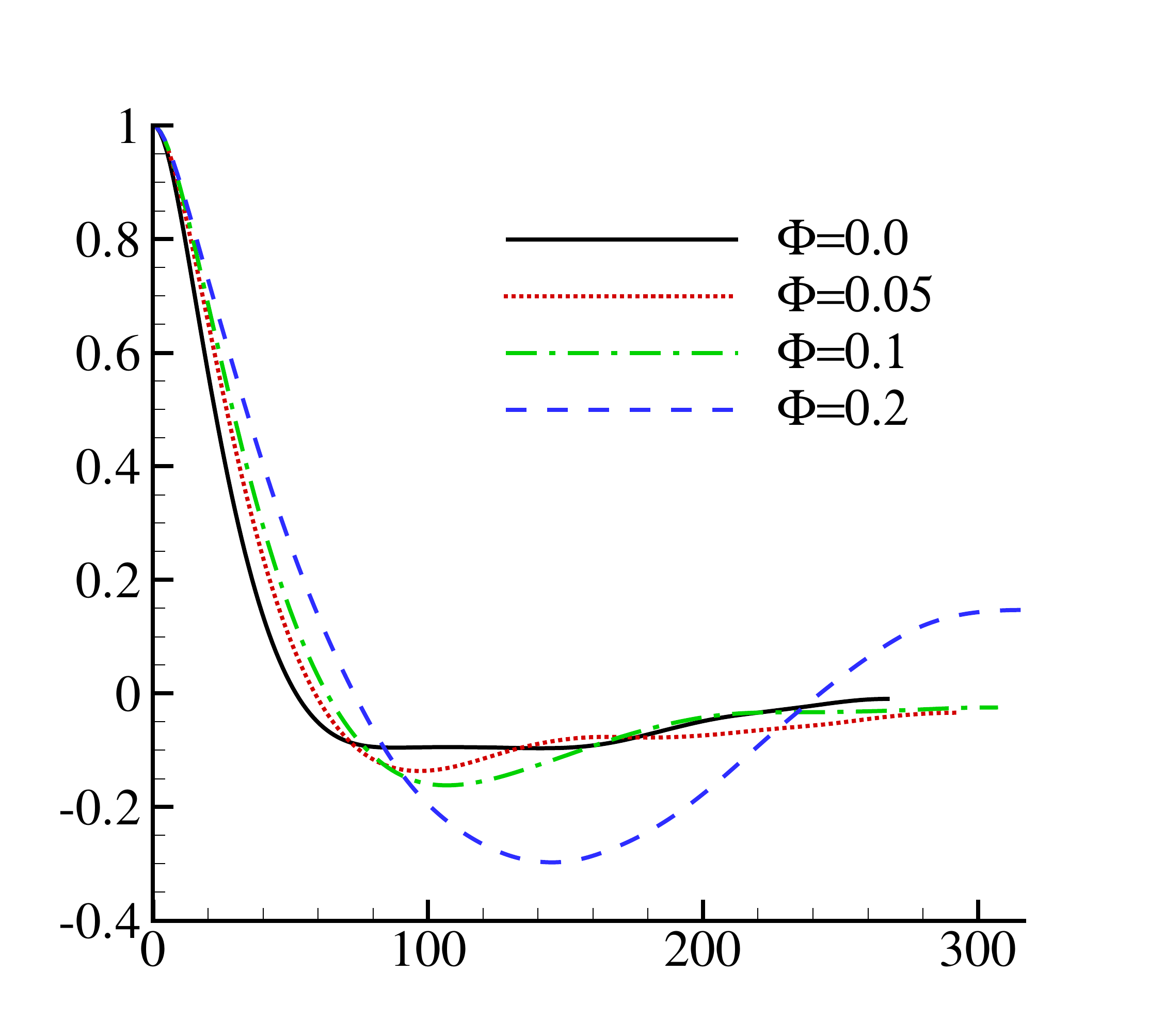} 
\put(-150,140){$b)\;y=2d$}\put(-100,2){$\Delta z^+$}\put(-188,75){\rotatebox{90}{$R_{uu}$}}\\[.1cm]
 \includegraphics[width=.49\linewidth]{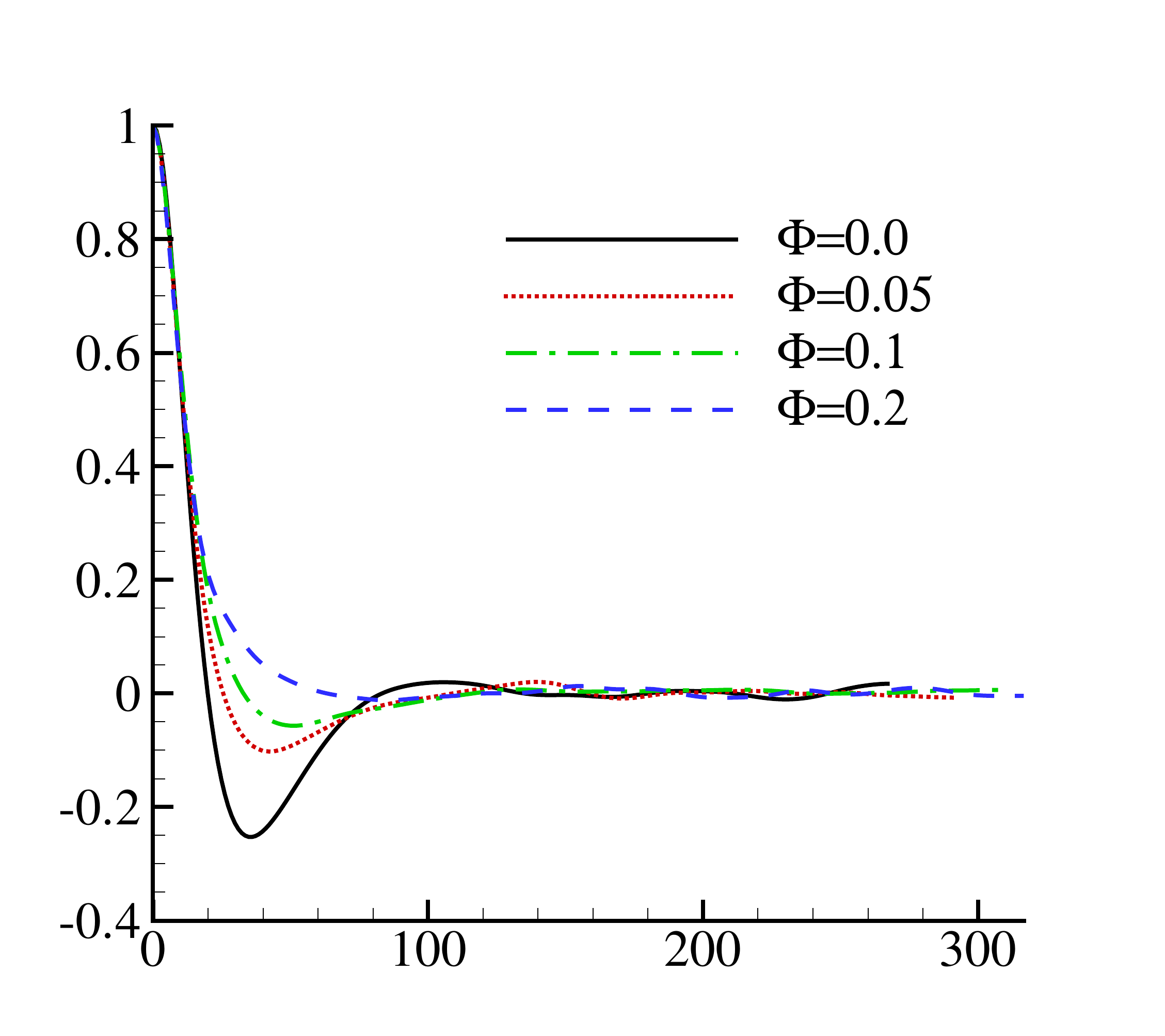} 
\put(-150,140){$c)\;y=d$}\put(-100,2){$\Delta z^+$}\put(-188,75){\rotatebox{90}{$R_{vv}$}}\hfill
 \includegraphics[width=.49\linewidth]{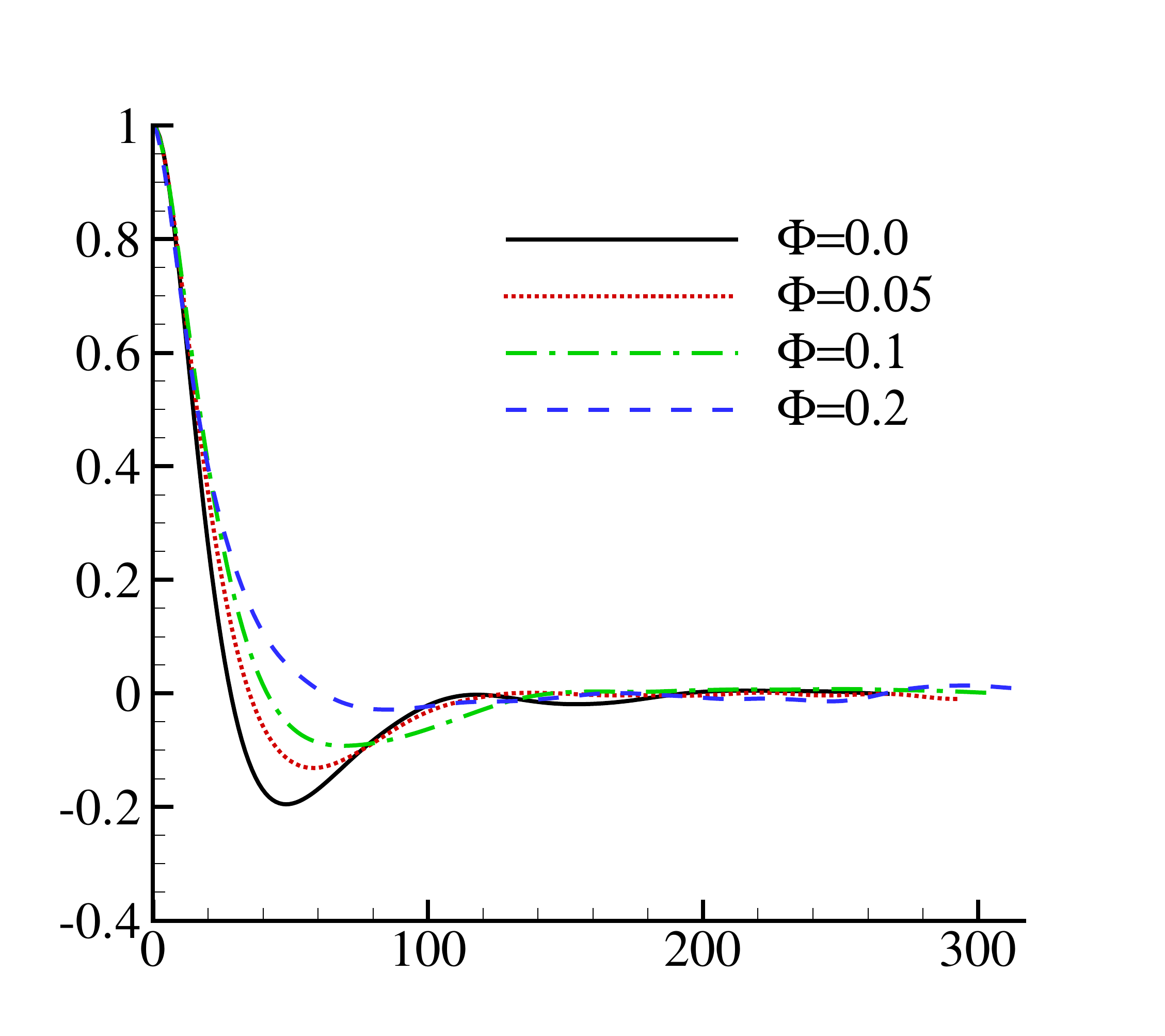} 
\put(-150,140){$d)\;y=2d$}\put(-100,2){$\Delta z^+$}\put(-188,75){\rotatebox{90}{$R_{vv}$}}
\caption{\label{fig:9} Correlations of the velocity fluctuations versus the spanwise separation $\Delta z^+$ for different bulk volume fractions $\Phi$.
Streamwise-streamwise component  $R_{uu}$ at $a)$: $y=d=h/18\simeq 20 \delta_*$  and $b)$ at 
$y=2d=h/9\simeq 40 \delta_*$.
Wall-normal component  $R_{vv}$ at $c)$ $y=d=h/18\simeq 20 \delta_*$, and $d)$ at 
$y=2d=h/9\simeq 40 \delta_*$.}
 \end{center}
 \end{figure} 

The streamwise auto-correlation $R_{uu}$ is shown in figure~\ref{fig:8}$a)$ and $b)$, where it is evaluated at 
two wall-normal distances, $y=d\simeq20\delta_*$ and $y=2d\simeq40\delta_*$. The correlations are
here calculated for the combined phase, but they do not differ appreciably if calculated only for the fluid phase. 
At $y=d$ we note a progressive
increase of the separation distance with the particle volume fraction together with a smoothening of the minimum, indicating a less evident width of the near-wall flow structures. Further away from the wall, 
$y=2d$, we observe the formation of wider streamwise velocity streaks for the flow with $\Phi=0.2$, with a separation
of the minimum of the auto-correlation, $\Delta z^+$, that is almost twice that pertaining to single-phase near-wall turbulence. 
The system tends therefore to form streaks twice as large as those in single-phase turbulent  channel flows. These larger
structures are also seen by the shift of the lowest minimum of $R_{uu}$ to $y=2d$ instead of $y=d\sim 20\delta_*$ 
where the single-phase channel flow shows the sharper minimum in the auto-correlation functions. 
The wall-normal auto-correlations $R_{vv}$ are shown in figure~\ref{fig:8}$c)$ and $d)$ for  the same two wall-parallel planes. 
Increasing the volume fraction $\Phi$ we observe less sharp minima that completely
disappears at $\Phi=0.2$. This suggests a significant alteration of the structure of the wall turbulence at high volume fractions
with a flow much less organised in coherent structures. Similar observations are reported in \cite{loisel2013effect} for transitional flows at lower $\Phi$.
The behaviour of the velocity auto-correlations is consistent with what found in turbulent drag reducing flows (growth of the buffer region). 
Hence it appears once more that despite the total drag increase, the turbulent induced drag reduces at least at high volume fraction.    

\section{Final remarks}

We report data from the numerical simulations of turbulent channel flow 
laden with finite-size particles at high volume fractions. The simulations have been performed using an efficient implementation of the Immersed Boundary Method that enable us to fully resolve the fluid-structure interactions.
We provide a statistical analysis to assess the effect of an increasing solid volume fraction (up to $\Phi=0.2$) on 
a turbulent channel flow at fixed bulk Reynolds number, i.e.\ $Re_b= U_0\, 2h /\nu$. 

The finite-size particles interact with the turbulent motions  altering the near-wall turbulence regeneration process. 
For the two lowest volume fractions considered, $\Phi \le 0.1$, we  still observe the classic behaviour of near-wall turbulence, modulated however by the particle presence.
At $\Phi=0.2$ the solid phase is so dense that several aspects of turbulent wall flows are
lost: the mean velocity profile is strongly altered, the turbulent fluctuations decrease, the velocity auto-correlations show streamwise elongated structures twice as wide 
as in single-phase channel flows and the absence of a negative correlation of the wall-normal velocity, in addition to a more isotropic distribution of the velocity fluctuations. 

The law of the wall is modified by the presence of a solid phase but can still be recognised at the Reynolds number of our simulations  for all the volume fractions investigated. The  von K{\'a}rm{\'a}n and additive constants, $k$ and $B$, assume therefore different values. In particular, increasing the volume fraction 
we report a reduction of $k$, increase of the slope, and a strong decrease of $B$, increased near-wall dissipation. 
The reduction of $k$ usually denotes turbulent drag reduction. However, in the present cases we always
observe an increase of the overall drag due to the decrease of the additive constant $B$. This 
is also confirmed by the increase of the friction Reynolds number, $Re_\tau$, when increasing the volume fraction at constant mass flux. 

We evaluate the streamwise momentum balance for the flows under investigations and show that the additional stress due to the presence of the particles becomes more and more relevant when increasing the particle volume fraction. As expected the Reynolds transport term dominates at zero and low $\Phi$, while at $\Phi=0.2$ the particle stress becomes of the same order of magnitude. 

Examining the turbulent shear stress and the streamwise momentum balance, we thus note that the
turbulence activity and the related stress reduce at the highest volume fraction here considered, i.e.\
$\Phi=0.2$. In order to characterise the \emph{turbulent} drag, we define a \emph{turbulent} friction
Reynolds number $Re_T$ whose friction velocity is based on the slope of the Reynolds shear stress
profile at the centreline. This parameter approximates the usual $Re_\tau$ in unladen turbulent channel flow. 
Using this turbulent friction Reynolds number, we quantitatively show that the \emph{turbulent} drag (measured by $Re_T$)
first gently increases with $\Phi$
and then sharply decreases at $\Phi=0.2$, even though the overall drag still increases.

These results suggest that further increasing the Reynolds number while keeping constant the particle size in inner units, $d^+$, may lead to an overall drag reduction in dense cases as those studied here. 
The main assumption behind this conjecture is that the near-wall turbulence-particle dynamics  remain similar when 
the bulk Reynolds number is increased, as it might occur when the particle size in inner units remain constant
(i.e.\ the friction particle Reynolds number).  
Indeed, we show here that increasing the bulk Reynolds number the turbulent induced drag increases its weight in
the stress balance. 
Hence, the reduced turbulence activity and the consequent reduced turbulent drag should induce
a decrease of the total drag at high enough Reynolds. This should appear as an extension of the log-layer with almost the same 
reduced $k$ and $B$ as reported here. 
New and even larger simulations would be needed in the future to test this hypothesis. 
In the meanwhile, we hope to stimulate new experimental investigations towards this direction. 

This study reports detailed statistics of particle laden channel flow at high volume fractions, accessible only recently \citep{kidanemariam2014direct,vowinckel2014fluid}, and it could be therefore extended in many non-trivial directions. Two-body particle statistics, such as collisions rates and clustering are not considered yet because out of the scope of the present 
work. In addition, the effect of the particle shape  \citep{bellani712shape} and deformability \cite[e.g.][]{clausen2011rheology} surely deserves attention as it will add new interesting physics to our current understanding.

\begin{acknowledgments}
This work was supported by the European Research Council Grant No.\ ERC-2013-CoG-616186, TRITOS.
The authors acknowledge computer time provided by SNIC (Swedish
National Infrastructure for Computing) and the support from the COST Action MP0806: \emph{Particles in turbulence}.
\end{acknowledgments}
 
\appendix

\section{Total stress of the suspension mixture}
\label{app}
In this work we use the framework developed by Prosperetti and co-workers to examine the stresses in suspension mixtures,
see e.g. \cite{mar_etal_ijmf99,zhapro_etal_pf10} for more details.

We assume the same density $\rho$ for the fluid and the particles and consider dimensional variables for all the calculation 
presented in this appendix. Following \cite{zhapro_etal_pf10}, we define  the phase indicator $\xi=0$ in the fluid phase and 1 in the solid one. 
{
Defining the phase-ensemble average, `$\langle \rangle$', as the ensemble average (implicitly) conditioned to the
phase considered (particulate, fluid and combined), we can calculate the local volume fraction in a point as}
\begin{equation}
\phi=\langle \xi \rangle.
\label{eq:phi_def}
\end{equation}
{
Considering a generic observable of the combined phase $o_c=\xi o_p+(1-\xi)o_f$, constructed in terms of 
$o_{p/f}$, the same observable in the particulate and fluid phases, it holds that
}
\begin{equation}
\langle o_c \rangle= \langle \xi o_p\rangle +\langle (1-\xi)o_f \rangle=  \phi \langle  o_p\rangle +(1-\phi) \langle o_f \rangle.
\label{eq:ens_average}
\end{equation}
{
Note that we are not using different symbols for the different phase ensemble averages, but implicitly assume that the phase conditioning is indicated by 
the sub-script inside the brackets.}  

The force balance for the volume $\cal V$ delimited by the surface $\cal S(V)$ is,

\begin{align}
\rho \int_{\cal V} \xi \mathbf{a}_p + (1-\xi) \mathbf{a}_f\, dV = \oint_{\cal S(V)} [\xi \pmb{\sigma}_p + (1-\xi) \pmb{\sigma}_f]
\cdot \mathbf{n} \,dS,
\label{eq:balance}  
\end{align}
 with $\mathbf{n}$ the outer unity vector normal to the surface  $\cal S(V)$, the subscripts `f' and `p' denoting fluid and particle phases, $\mathbf{a}_i$ and $\pmb{\sigma}_i$ the acceleration and the general stress in the phase $i$. 
Applying the phase ensemble average to equation~\eqref{eq:balance},  
we obtain
\begin{align}
\rho \int_{\cal V} \langle \xi \mathbf{a}_p\rangle  + \langle (1-\xi) \mathbf{a}_f\rangle \, dV = \int_{\cal V} \nabla \cdot [\langle 
\xi \pmb{\sigma}_p\rangle  + \langle (1-\xi) \pmb{\sigma}_f\rangle] \,dV,
\label{eq:balance_ave}  
\end{align}
 where we  used the divergence theorem to the differentiable integrand on the right hand side. Since last equation 
 holds for any mesoscale volume $\cal V$, we can use the corresponding differential form of the equation,
\begin{align}
\rho \langle \xi \mathbf{a}_p\rangle  + \rho \langle (1-\xi) \mathbf{a}_f\rangle  = \nabla \cdot [\langle 
\xi \pmb{\sigma}_p\rangle  + \langle (1-\xi) \pmb{\sigma}_f\rangle].
\label{eq:balance_ave_diff}  
\end{align}   
Considering the identities \eqref{eq:phi_def}-\eqref{eq:ens_average}, we can further simplify the expression above

\begin{align}
\rho \phi \langle \mathbf{a}_p\rangle  + \rho (1-\phi) \langle  \mathbf{a}_f\rangle  = \nabla \cdot (\phi \langle 
 \pmb{\sigma}_p \rangle  + (1-\phi) \langle \pmb{\sigma}_f\rangle).
\label{eq:balance_ave_diff_phi}  
\end{align}   
Assuming the constitutive law of a Newtonian fluid 
$\pmb{\sigma}_f=-p \mathbf{I} +2\mu \mathbf{E}$ with $p$ the pressure and $\mathbf{E}=(\nabla \mathbf{u}_f+
{\nabla \mathbf{u}_f}^T)/2$ 
the symmetric part of
the fluid velocity gradient tensor and considering that both the fluid and particle velocity fields are divergence-free, eq.~\eqref{eq:balance_ave_diff_phi} can be re-written as,
\begin{align}
\phi \frac{ \langle \mathbf{u}_p \rangle}{\partial t} &+  \phi  \langle {\mathbf{u}_p}\cdot\nabla{\mathbf{u}_p}  \rangle +  (1-\phi)  \frac{\langle\mathbf{u}_f\rangle}{\partial t} +  (1-\phi) \langle {\mathbf{u}_f}\cdot\nabla{\mathbf{u}_f}    \rangle  =\nonumber \\ 
&\nabla \cdot \big(\phi \langle 
{\pmb{\sigma}_p}/{\rho} \rangle\big)  - 
{ \nabla \big[ (1-\phi) \langle p/{\rho}\rangle\big]} +
\nabla\cdot \big[ (1-\phi)2\nu\langle \mathbf{E} \rangle\big].
\label{eq:balance_ave_diff_phi2}  
\end{align}
We next denote the statistically stationary mean fluid and particle velocities as $\mathbf{U}_{f/p}= \langle \mathbf{u}_{f/p} \rangle$ and the
fluctuations around these mean values as $\mathbf{u'}_{f/p}=\mathbf{U}_{f/p}- \langle  \mathbf{u}_{f/p} \rangle$, so that  the average momentum equation becomes

\begin{align}
\phi  {\mathbf{U}_p}\cdot\nabla{\mathbf{U}_p} +&\phi  \nabla \cdot \langle {\mathbf{u'}_p} {\mathbf{u'}_p}  \rangle  +  (1-\phi)  {\mathbf{U}_f}\cdot\nabla{\mathbf{U}_f}     +
(1-\phi) \langle {\mathbf{u'}_f} {\mathbf{u'}_f}    \rangle =\nonumber \\ 
&\nabla \cdot \big(\phi  
{\langle\pmb{\sigma}_p/{\rho}\rangle}\big)-  
{ \nabla \big[ (1-\phi) P/{\rho}\big]} +
\nabla\cdot \big[ (1-\phi)2\nu\langle \mathbf{E} \rangle\big],
\label{eq:balance_ave_diff_phi_mean}  
\end{align}
\noindent with $P$ the mean pressure.

Exploiting the symmetries of a fully developed parallel channel flow, characterised by two homogeneous directions --the streamwise, $x$  and spanwise,
$z$--, we  project eq.~\eqref{eq:balance_ave_diff_phi_mean} 
in the inhomogeneous wall-normal direction $y$,
  \begin{align}
\frac{d}{d y}  \left[(1-\phi)\langle v'^{2}_f\rangle + \phi\langle v'^{2}_p\rangle + (1-\phi)\frac{P}{\rho} - \frac{\phi}{\rho}\langle\sigma_{p\,yy}\rangle\right]= 0.
\label{eq:momentum_y1}
\end{align}
 Integrating equation (\ref{eq:momentum_y1}) in the $y$--direction and denoting by $P_w(x)$ the wall pressure, we obtain
\begin{align}
(1-\phi)\langle v'^{2}_f\rangle + \phi\langle v'^{2}_p\rangle + \frac{P_T}{\rho}= \frac{P_w}{\rho},
\label{eq:momentum_y2}
\end{align}
where we also introduced the mean total pressure $P_T=(1-\phi)({P}/{\rho}) - {\phi}\langle \sigma_{P\,yy}\rangle/\rho$. 
It should be noted that $P_T$ coincides with $P_w$ at the wall and that
\begin{align}
\frac{\partial P_T}{\partial x}=\frac{\partial P_w}{\partial x}.
\label{eq:tot_press}
\end{align}

Projecting eq.~\eqref{eq:balance_ave_diff_phi_mean} in the streamwise direction $x$, we have
\begin{align}
\frac{d}{d y}  \left[(1-\phi)\langle u'_f v'_f \rangle + \phi \langle u'_p v'_{p}\rangle - \nu(1-\phi)\frac{d U_f}{d y} - \frac{\phi}{\rho}\langle\sigma_{p\,xy}\rangle\right]=-\frac{d}{d x} (\frac{P_w}{\rho}),
\label{eq:momentum_x3}
\end{align}
where we neglect the terms $({\partial}/{\partial x}) \big{[}({\phi}/{\rho}) (\langle \sigma_{p\,xx} - \sigma_{p\,yy}\rangle)\big{]}$ because of the
streamwise homogeneity.

Integrating eq.~\eqref{eq:momentum_x3} in the wall normal direction and denoting the Reynolds shear stress of the combined
phase $\langle u'_{C}v'_{C}\rangle=(1-\phi)\langle u'_f v'_f \rangle + \phi \langle u'_p v'_{p}\rangle$, we obtain the equation for the total stress $\tau(y)$, 
\begin{align}
\tau(y)= - \langle u'_{C}v'_{C}\rangle + \nu(1-\phi)\frac{d U_f}{d y} + \frac{\phi}{\rho}\langle \sigma_{p\,xy}\rangle=
\nu \frac{d U_f}{d y}\big{|_w} (1-\frac{y}{h}),
\label{eq:total_stress}
\end{align}
where we considered the boundary condition at the wall, $\tau_w=\tau(0)=\nu ({d U_f}/{d y})|_{y=0}$.
Eq.~\eqref{eq:total_stress} shows that the total stress of a turbulent suspension in a channel geometry is given
by three contributions: the viscous part, $\tau_V= \nu(1-\phi)({d U_f}/{d y})$, the turbulent part $\tau_T=-\langle u'_{C}v'_{C}\rangle=-
(1-\phi)\langle u'_f v'_f \rangle - \phi \langle u'_p v'_{p}\rangle$ and the particle-induced stress, $\tau_P={\phi}\langle \sigma_{p\,xy}
\rangle/{\rho}$. It should be noted that the turbulent stress accounts for the coherent streamwise and wall-normal motion of both
 fluid and solid phases. The particle induced stress is originated by the total stress exerted by the solid phase, 
 see eq.~\eqref{eq:balance}, and takes into account hydrodynamic interactions and collisions.   
 In the absence of particles, $\phi \to 0$, eq.~\eqref{eq:total_stress} reduces to the classic momentum balance for single phase turbulence \citep[see][]{pope2000turbulent}.

\bibliographystyle{jfm}

\bibliography{jfm}

\end{document}